\documentclass[final,5p,times,twocolumn]{elsarticle}

\usepackage{amssymb}
\usepackage{amsthm}
\usepackage{amsmath}
\usepackage{cases}
\usepackage{graphics}
\usepackage{natbib}
\begin{document}

\begin{frontmatter}



\title{Development and Virtual Testing of 5G Connected Adaptive Cruise Control}


\author[inst1]{Andrea Leo}

\affiliation[inst1]{organization={Department of Mechanical Engineering},%Department and Organization
            addressline={Via La Masa, 1}, 
            city={Milan},
            postcode={20156}, 
            %state={Italy},
            country={Italy}}

\author[inst1]{Giacomo Reatti}
\author[inst1]{Stefano Arrigoni}
\author[inst1]{Michael Khayyat}
\author[inst1]{Federico Cheli}

%\affiliation[inst2]{organization={Department Two},%Department and Organization
%            addressline={Address Two}, 
%            city={City Two},
%            postcode={22222}, 
%            state={State Two},
%            country={Country Two}}

\begin{abstract}
It is estimated that 90\% of crashes occur due to human error, mainly induced by poor judgement, distraction or lack of situation awareness \cite{peters2002automotive}.
The development of systems that aim to increase road safety and to improve driving comfort is the key challenge that many OEMs are currently tackling. The use of fast and reliable of innovative communication technologies such as 5G can enhance advancements and developments in the field of ADAS. 
The aim of the paper is the study of the potential improvements connectivity can bring to ACC systems in terms of safety and vehicle comfort. Starting from state-of-art ACC present in literature, a novel connected ACC is designed an deeply reported. The main features of the algorithms consist in integrating information about other road vehicles fed by 5G and about potential grip given by smart sensors such as Pirelli Cyber Tyre. The novel solution is then compared to the standard ACC by means of several simulations considering two real traffic situations with different road friction coefficients. Finally benefits and limitations of the proposed algorithm are deeply discussed.
%A base case, commercial ACC system is developed. Then, a novel Connected ACC algorithm is proposed. It is an improvement of the existing commercial ACC logic: information about other road vehicles fed by 5G and about potential grip given by smart sensors such as Pirelli Cyber Tyre are utilized to produce a CACC.
%The developed CACC logic is tested and compared to the commercial ACC in a virtual environment in two real traffic situations with different road friction coefficients. Simulation-based results show the high potential of connectivity in increasing traffic safety and comfort.
\end{abstract}



\end{frontmatter}


\section{Introduction}
\label{intro}
According to the World Health Organization (WHO), road traffic injury represents the eight cause of death in the world and the Leading cause of death for children and young adults aged 5-29 years \cite{WHO}. 
Every year the number of road deaths is around 1.3 million and between 20 and 50 million people sustain non-fatal injuries. Among the various causes, human errors seem to be the main cause of road accidents. 
In fact, speeding contributes up to 35\% of fatal road crashes and driving under the influence of alcohol contributes up to 30\% of them \cite{IFT}. Scientific research and car manufacturers are therefore focused on increasing road safety by introducing Advanced Driver Assistance Systems (ADAS). By adjusting human driver decisions and actions, ADAS aim to eliminate most of driver errors
and achieve a more regular and smooth vehicle control \cite{piao2008advanced}.
Recently, the evolution of ADAS is progressively shifting towards connected driving \cite{6834940}.
In particular, the use of 5G for V2V and V2I transmissions is gaining importance due to its reliability, speed and ability to support a high demand of data.
Features like Adaptive Cruise Control (ACC) were introduced to increase driving comfort and traffic safety, especially during the use in highways or non-urban roads. The evolution into a Connected ACC permits to have an improved behaviour in dangerous situations, especially in ones involving side and rear-end collisions,\cite{validi2017analyzing},
and to reduce traffic jams yet maintaining a high traffic throughput \cite{ploeg2011connect}. Hence, the development and the testing of a Connected Adaptive Cruise Control systems that use 5G V2V communication system and Pirelli Cyber Tyre to estimate potential grip is therefore proposed.
%\appendix

\section{Literature Review}
\label{LitRev}
\subsection{Adaptive Cruise Control Systems}
The Adaptive Cruise Control (ACC) is a technology that extends the capabilities of the standard Cruise Control (CC) \cite{rajamani2006vehicles}. Cruise Control is capable of making the vehicle travel at a user-set speed, typically selected through steering wheel commands. ACC is supported by range sensors that make the car aware of the condition of the Preceding vehicle and therefore can adjust the speed according to the relative distance and velocity with it.

ACC has been introduced in commercial vehicles in Japan since 1995, followed by Europe since 1998 and North America since 2000 \cite{xiao2010comprehensive}.
They were firstly introduced as devices aimed at increasing passengers driving comfort. However, soon researchers and car companies started considering them also as systems able to deal with safety issues and improvement of traffic congestion.
 
\subsubsection{System Architecture}\label{ACC_sys_arch}

\begin{figure}
\centering
\includegraphics[scale=0.5]{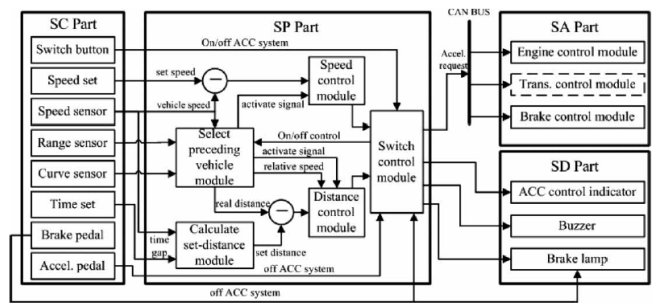}
\caption[Example of ACC system architecture]{Example of ACC system architecture \cite{xiao2010comprehensive}}
\label{ACC_sys_archit}
\end{figure}

The typical architecture of a ACC is represented in Figure \ref{ACC_sys_archit}.
Here four main modules can be recognized: 
\begin{enumerate}
    \item \textbf{Signal Collecting (SC)}: this module is responsible for providing the whole logic the necessary inputs. 
    Relative distance and relative speed are obtained through the radar sensor, while other sensors are able to detect the path that the car is going to approach. 
    \item \textbf{Signal Processing (SP)}: in this part the control logic of the ACC system is located. The picture shows a typical control logic structured into a Speed Control module and a Spacing Control one (the latter indicated as \textit{Distance Control module}).
    \item \textbf{Signal Actuating (SA)}: the desired acceleration computed in SP module is here converted to throttle or brake command. If the vehicle is provided with a gearbox, also a transmission control module is needed to select the correct gear.
    \item \textbf{Signal Displaying (SD)}: when the ACC is activated, the control indicator displays the operating state (speed control or spacing control). 
\end{enumerate}

Another scheme usually considered to specifically describe controller logic architecture is reported in \cite{rajamani2006vehicles} and shown in Figure \ref{lowhighlevel}:

\begin{figure}
    \centering
    \includegraphics[scale=0.28]{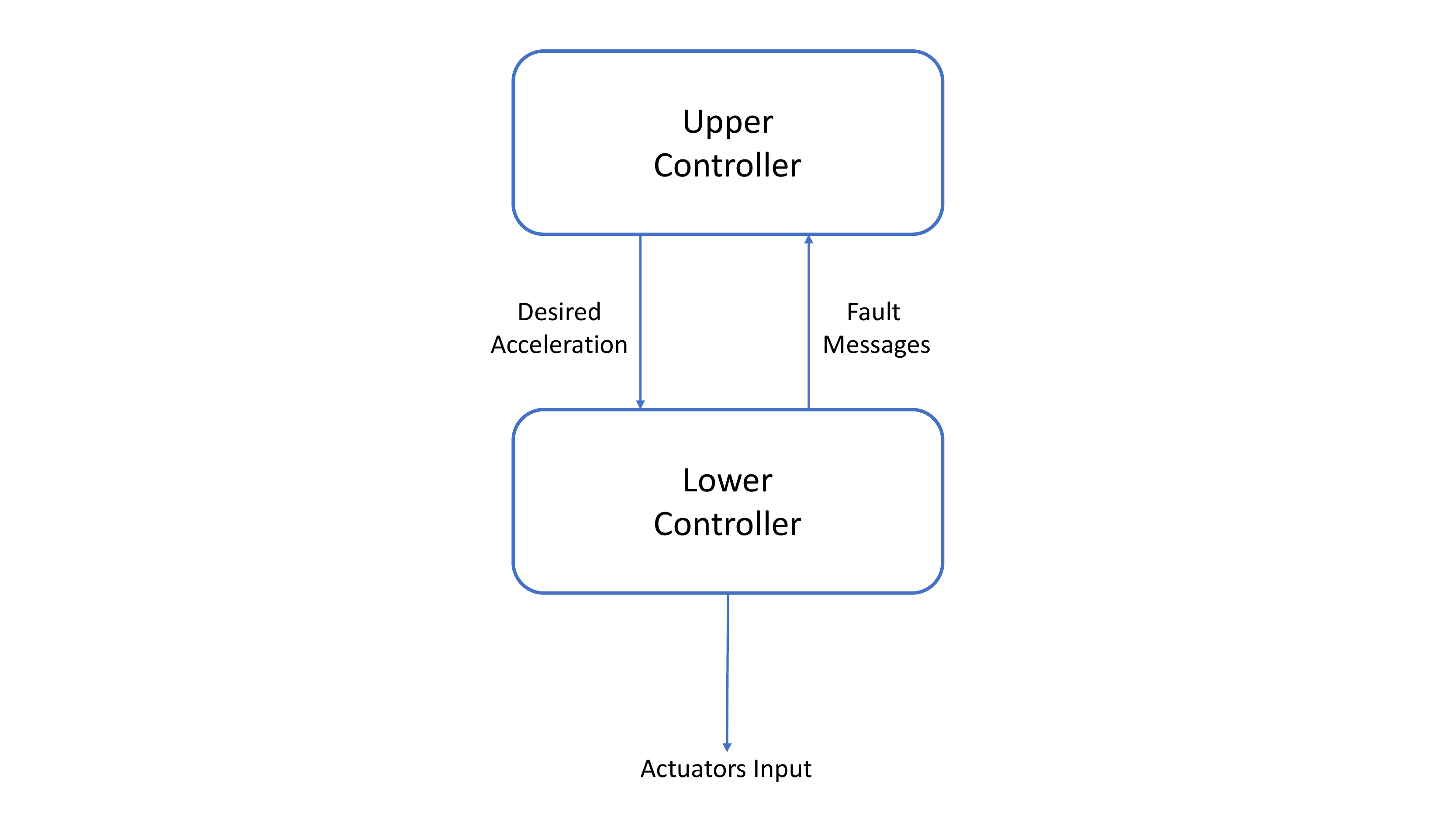}
    \caption{Upper and Lower Level Scheme}
    \label{lowhighlevel}
\end{figure}

ACC control algorithm is designed through a hierarchical scheme composed by two levels. The upper level controller is properly called ACC controller, while the lower level is called longitudinal controller. The output of the ACC controller is the desired acceleration which is transmitted to the longitudinal controller. The latter determines the throttle and/or brake command using vehicle dynamic models, engine maps and control techniques.

The desired acceleration is limited in a range of values. The choice of the extremes of the range is, in fact, a design parameter. In Table \ref{accel_decel_ACC}
the values used by literature for the maximum acceleration and deceleration are reported.

\begin{table}
\footnotesize %necessary command to reduce font size for the table. If problems arises, find a new command
    \centering
     \begin{tabular}{c c c} 
     \hline
     Reference & Max. Acceleration [$m/s^2$] & Max. Deceleration [$m/s^2$]\\
     \hline
     \cite{li2010model} & 0.6 & -1.5 \\ 
     
     \cite{prestl2000bmw} & 1 & -2 \\
    
     \cite{winner1996adaptive,bosch2003acc} & 1 & -2.5 \\
    
     \cite{rajamani2006vehicles,wang2002adaptive,wang2004should} & 2 & -5 \\
     \hline
    \end{tabular}
    \caption{Acceleration and deceleration limits for ACC systems according to literature}
    \label{accel_decel_ACC}
\end{table}
Apart from acceleration constraints, some ACC systems consider acceleration rate (also referred to as jerk) as part of the optimization process when high importance is given to driving comfort \cite{naus2010model,li2010model}. 
Reference limitation values are 0.1 $m/s^3$ for positive jerk and -1 $m/s^3$ for negative jerk.
Another constraint is that ACC automatically deactivates when the speed is lower then a certain threshold (typically of $25-40 km/h$). As discussed in \cite{sensorsautomotive}, ACC systems find their typical application in highways, while their introduction into urban traffic requires the use of complicated additional sensors to properly recognize all objects and obstructions. Moreover, automatic positive acceleration at low speed has to be avoided for safety reasons \cite{winner1996adaptive}.

Most of the developed ACC systems have two modes of operation:
\begin{itemize}
\setlength\itemsep{0em}
    \item Speed control (often referred to as \textit{Velocity Control})
    \item Spacing control (or \textit{Vehicle Following}, \textit{Distance Control})
\end{itemize}
In order to shift from one logic to the other, a switching method is needed. 
The purpose and need of a transitional control algorithm is to allow a smooth transition between these two operation modes and a reaction in case of a new preceding vehicle is taken as reference \cite{rajamani2006vehicles}. This situation is common in case of vehicle cut-in\footnote{A cut-in is defined as a manoeuvre where a vehicle travelling on lane suddenly moves to an adjacent one in between two other vehicles.} or lane change maneuver.

Different spacing control systems have been proposed. Linear controllers are the most common one for their programming simplicity and consists of a Proportional Integral Derivative controllers (PID)\cite{rajamani2006vehicles,martinez2007safe,wang2002adaptive,caudill2017vehicle,zhou2004string,pananurak2009adaptive,yanakiev1998nonlinear}. Their use for ACC systems is widespread in car companies, as many patents show \cite{winner1995patent,labhun1995patent}. Model Predictive Control (MPC) has been used more recently since it is able to solve at a higher computational cost multi-objective problems \cite{li2010model,bageshwar2004model,magdici2017adaptive,luo2010model}. Moreover, fuzzy and neural controllers have been suggested to replicate a car-following behavior similar to the humans would have \cite{naranjo2006acc,naranjo2003adaptive,ko2007fuzzy,fancher1996comparative}. Finally, also the sliding surface method derived from  \cite{slotine1991applied} has been used in Rajamani \cite{rajamani2000design}, providing accurate and reliable performances.
%ci sarebbe engelman2001patent da citare, ma da un errore che non mi spiego
\subsubsection{Spacing Policies}

The definition of a ACC system starts with the definition of a spacing policy \cite{swaroop2001review}. When the controlled vehicle operates in Spacing Control, a certain spacing policy has to be selected, which is responsible for defining the desired steady state distance between two successive vehicles. The choice of a suitable spacing policies is fundamental both for safety reasons since the time available to brake in emergency scenarios is determined by the inter-vehicle distance and for comfort reasons since a too aggressive policy can cause unnecessary acceleration and deceleration which lead to a low comfort for the passengers.
\begin{figure}
        \centering
            \includegraphics[scale=0.45]{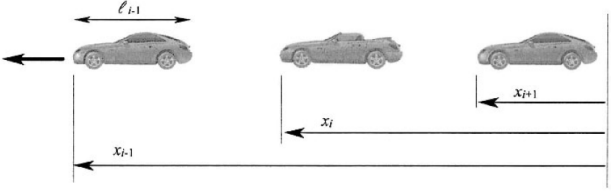}
             \caption[Notation for spacing policies]{Notation for spacing policies \cite{rajamani2006vehicles}}
             \label{raja_notation}
    \end{figure}
\noindent Many spacing policies are present in literature and discussions are still open on which one can guarantee better performances. 
In the following, the index notation of Figure \ref{raja_notation} 
will be used. Among the numerous policies, the most widespread are:

\begin{itemize}
    \item \textbf{Constant distance spacing (CD)}:
    \newline In this kind of policy, two consecutive vehicles must keep a constant distance $L_{desired}$. The inter-vehicle distance is defined in equation \ref{interv_spac_CD}:
    \begin{equation}
        \epsilon_{i} = x_{i} - x_{i-1} + l_{i-1} \label{interv_spac_CD}
    \end{equation}
    where $x_{i}$ and $x_{i-1}$ are the position of the two vehicles and $l_{i-1}$ the length of the Preceding vehicle. The spacing error of the $i-th$ vehicle is then defined in equation \ref{spacing_err_CD}:
    \begin{equation}
        e_{p,i} = x_{i} - x_{i-1} + L_{desired} \label{spacing_err_CD}
    \end{equation}
    Assuming that the acceleration of the vehicle can be controlled immediately, a linear controller is applied, as shown in \ref{controller_CD}:
    \begin{equation}
        \Ddot{x}_{i} = -k_{p}e_{p,i} -k_{v}\Dot{e}_{p,i} \label{controller_CD}
    \end{equation}
    where $k_{p}$ and $k_{v}$ are control gains.
    
    With this kind of controller, Rajamani \cite{rajamani2006vehicles} demonstrates that a linear control system that adopts the constant spacing policy cannot guarantee string stability for autonomous vehicles that are not provided with inter-vehicle communication.
    
    \item \textbf{Constant time gap spacing (CTG)}:
    \newline In CTG policy, the inter-vehicle distance is not constant. As shown in \ref{interv_spac_CTH}, it varies linearly with the following vehicle velocity:
    \begin{equation}
        L_{desired} = d_{min} + h\Dot{x}_{i} \label{interv_spac_CTH}
    \end{equation}
    where $h$ is the desired time gap and $d_{min}$ is the inter-vehicle distance at rest. Time gap is defined as the time it takes for a vehicle to cover the distance measured from the front of the following vehicle to the rear of the preceding one \cite{bose2001evaluation,bose2001analysis}. Consequently, the spacing error varies linearly with velocity, as represented in \ref{spacing_err_CTH}:
    \begin{equation}
        e_{p,i} = \epsilon_{i} + h\Dot{x}_{i} \label{spacing_err_CTH}
    \end{equation}
    where $\epsilon_{i}$ has the same expression of \ref{interv_spac_CD}.
    Iannou and Chien \cite{ioannou1993autonomous} proposed the control law in \ref{controller_CTH}:
    \begin{equation}
        \Ddot{x}_{i,des} = -\dfrac{1}{h}(\Dot{\epsilon_{i}}+\lambda e_{p,i})  \label{controller_CTH}
    \end{equation}
    where $\lambda$ is a control parameter and has to be chosen greater then zero \cite{swaroop1994comparision}. Again, computing the transfer function between spacing error of following vehicle and spacing policy of Preceding vehicle, its magnitude is always equal or less then the unity only if the necessary and sufficient condition represented in \ref{condition_CTH} 
    is respected:
    \begin{equation}
        h \geq 2\tau \label{condition_CTH}
    \end{equation}
    where $\tau$ is the time constant of the lower level control (typically its value is around 0.5$s$). Therefore, in general time gap should be increased over 1$s$ to avoid string stability: this represents a limit for this kind of logic when the aim is to maximise the traffic capacity.
    
    %Swaroop et al.\cite{darbha1999intelligent} demonstrate that when all the vehicles that flow on a highway adopt a CTG policy, then traffic flow is unstable. Traffic flow in this case is defined as the increasing upstream transmission of disturbances that occur when traffic condition are modified from its steady state condition.
    
    %Hedrick et al.\cite{swaroop1994comparision} perform a comparison between CD and CTG policies using the lane capacity in $vehicles/lane/hour$ as a measure. Assuming a constant distance of 1$m$, a time gap of 0.2$s$ and different platoon sizes, results show that CD policy has a 25\% more traffic capacity then CTG. However, as already mentioned, string stability becomes an issues with this policy and a safety analysis is not considered.
    
    \item \textbf{Variable time gap spacing (VTG)}:
    \newline An evolution of CTG policy for ACC systems has been made by Santhanakrishnan \cite{santhanakrishnan2003spacing} and Wang \cite{wang2002adaptive}. It aims at satisfying string stability and traffic flow stability, still guaranteeing high traffic flow capacity. The policy has a nonlinear dependency on speed unlike the common one. Starting from the spacing policy shown in \ref{spac_err_CTH_nonlin}:
    \begin{equation}
        e_{p,i} = \epsilon_{i} + g(\Dot{x}_{i}) \label{spac_err_CTH_nonlin}
    \end{equation}
    where $g(\dot{x}_{i})$ is the nonlinear function that substitutes the linear term with time gap of \ref{spacing_err_CTH}.
    By linearizing this policy, it is possible to get a local condition for string stability, as shown in \ref{spac_err_CTH_nonlin_ss}:
    \begin{equation}
        \frac{\partial g}{\partial \dot{x}_{i}} > 2\tau \label{spac_err_CTH_nonlin_ss}
    \end{equation}
    This condition is then evaluated in term of traffic volume rate $Q$ and highway vehicle density $\rho$ in order to add the requirements of traffic flow stability, which is $\frac{\partial Q}{\partial \rho} > 0$. However, this is possible only for certain values of $\rho$: in fact, the traffic flow rate drops to zero at maximum density, so that a negative slope is inevitable. %(also see %\ref{traff_flow_ins_section}).
    Therefore these results are restricted only to particular situations and the control designer should be aware in which scenarios its logic will be mainly used.\newline
    %In \cite{wang2002adaptive} a similar logic is proposed and also safety issues are taken into consideration since a higher traffic capacity means smaller inter-vehicle distances. The following scenarios with 8 vehicles are analyzed:
   % \begin{itemize}
    %\setlength\itemsep{0em}
    %    \item Sudden braking of Preceding vehicle of 5 $m/s^2$;
    %    \item ACC vehicle detects a vehicle which is 10 $m/s$ slower;
    %    \item Vehicle cuts into ACC vehicle at short distance 5 $m/s$ %slower.
    %\end{itemize}
    
    %In the first scenario, the VTG policy allows keep a higher inter-vehicle spacing during the transient braking with respect to the CTG policy.\newline
    %In the second scenario, VTG and CTG performs in a similar way, both guaranteeing a smooth transition to the steady state behaviour.\newline
    %In the last scenario, the safety performance of the VTG system is better than that of the CTG system.
    \item \textbf{Variable separation error gain spacing}:\newline
    This logic \cite{yanakiev1998nonlinear} finds interesting application in heavy-duty vehicles that require larger inter-vehicle distances due to the higher difficulties for their actuators to act on such a big inertia. Considering spacing error (\ref{delta_VSEG})
    and relative speed (\ref{vrel_VSEG}):
    \begin{equation}
        e_{p,i} = (x_{i}-x_{i-1}) - (d_{min} + h\Dot{x}_{i}) \label{delta_VSEG}
    \end{equation}
    \begin{equation}
        v_{rel} = v_{prec} - v_{follow}  \label{vrel_VSEG}
    \end{equation}
    Putting together \ref{delta_VSEG} and \ref{vrel_VSEG} the control objective is shown in \ref{contr_obj_VSEG}:
    \begin{equation}
        v_{rel} + ke_{p,i} = 0  \label{contr_obj_VSEG}
    \end{equation}
    The idea is to vary the separation error gain $k$. In fact, if two vehicles are closer then desired ($e_{p,i}<0$) and the following vehicle is slower then the Preceding one ($v_{rel}>0$), the controller of the following vehicle does not need to react drastically. The same holds also if the vehicle are separating ($e_{p,i}>0$) but ($v_{rel}<0$): the vehicle behind should behave so that spacing error decreases smoothly.\newline
    %Therefore $k$, which is represented in \ref{/k_VSEG}, 
    $K$ is computed in \ref{k_VSEG_ph} as:
    \begin{equation}
        k = c_{k}+(k_{0}-c_{k})e^{-\sigma e_{p}^2}  \label{k_VSEG_ph}
    \end{equation}
    where $0<c_{k}<k_{0}$ and $\sigma\geq0$ are design constraints. According to which value of $k$ is chosen, the control objective becomes nonlinear in $e_{p}$. Simulation results show that the variable separation error gain spacing policy should be preferred when control smoothness and control robustness a priorities in control design.
    %\begin{figure}[h!]
    %    \begin{center}
    %        \includegraphics[scale=0.6]{pictures/State_of_art/k_VSEG.png}
    %          \caption[Separation error gain as function of $e_{p}$]{Separation error gain as function of $e_{p}$ \cite{yanakiev1998nonlinear}}
    %          \label{/k_VSEG}
    %    \end{center}
    %\end{figure}
    \item \textbf{User acceptance spacing}:\newline
    In \cite{han2006driver}, Han and Yi suggest a spacing policy that can accurately replicate the behaviour of a human driver. A recursive least-square algorithm has been applied to data collected from 125 human drivers to estimate driver tendency parameters. In particular, these variables have been selected:
    \begin{itemize}
    \setlength\itemsep{0em}
        \item Time gap
        \item Time to Collision (TTC), where:
        \begin{equation}
            TTC = \frac{x_{i}-x_{i-1}}{v_{i} - v_{i-1}}  \label{TTC}
        \end{equation}
    \end{itemize}
    The adopted controller is a quadratic range policy and it has been set according to these estimated parameters. The comparison between data from manual driving and with ACC system with this logic show a good similarity, showing that it is possible to obtain a driving behaviour that can prioritize user-acceptance. Similarly, in \cite{fancher2003research} a study has been performed on 107 human-driven vehicles to obtain an analytical expression of their range behaviour. Once more, a quadratic range policy has been proposed, as shown in \ref{human_dr_quadr}:
    \begin{equation}
        R = A + Tv_{i} + Gv^2_{i}  \label{human_dr_quadr}
    \end{equation}
    where A is the distance at standstill and T and G are coefficients found by curve fitting. The relationship between T and G is found to be almost linear. % and it is shown in \ref{curve_fitting_human_dr}. 
    Furthermore, negative values of G show that at high speed human drivers tend to reduce the time gap.
    %These results were used by \cite{zhou2005range}.
    %\begin{figure}[h!]
    %\begin{center}
    %    \includegraphics[scale=0.40]{pictures/State_of_art/curve_fitting_human_dr.png}
    %      \caption[Relationship between coefficients T and G]{Relationship between coefficients T and G \cite{fancher2003research}}
    %      \label{curve_fitting_human_dr}
    %\end{center}
%\end{figure}
\end{itemize}
\noindent Among all the aforementioned systems present in literature, from the point of view of the automotive industry, the most widespread adopted spacing policy results to be the CTG spacing policy, considering a good balance between safety, stability, reliability, feasibility and capability.

\subsection{Cooperative Adaptive Cruise Control}

Cooperative Adaptive Cruise Control (CACC) is defined as an ACC system equipped with V2V communication \cite{naus2010string, oncu2014cooperative,  milanes2013cooperative, dey2015review, van2006impact, lidstrom2012modular, naus2009towards}: with this kind of communication the vehicle gets information not only from its Preceding vehicle, as occurs in ACC, but also from the vehicles in front of the Preceding one \cite{milanes2013cooperative,lidstrom2012modular}. 
In this way, the controller is capable of better anticipate of risky situations thanks to the information of vehicles that are behind the so-called LOS (Line Of Sight), thus giving a smoother response in terms of driver comfort \cite{van2006impact} (mitigation of shock waves effects) and safer response in terms of road safety.
It is therefore important to notice that the CACC system has a potential beneficial effect on traffic efficiency and safety: the idea is to significantly reduce the time-gap to increase the traffic throughput \cite{ploeg2011design} with respect to the allowed ACC time-gap which is lower bounded by the string-stability requirements \cite{rajamani2006vehicles}.

\subsubsection{Control Strategies}
Many authors in literature focus on developing their CACC control logic by extending the functionality of an ACC system adding communication between vehicle and infrastructures. In particular, Dey et al.\cite{dey2015review} review literature on this aspect and they pointed out that, considering a string of vehicles, the majority of the CACC systems extensively exploits V2V communication with just the immediate nearest vehicle, contrary to communication with multiple Preceding vehicles or with a designated platoon leader. The change in type of vehicles communication is therefore reflected in different control strategies.

This kind of systems are generally referred to as Semi-Autonomous Adaptive Cruise Control (SAACC) and they have the advantage of being easy to be implemented.
%%RAJA
As a first example, Rajamani et al. \cite{rajamani2002semi} claim that the positive aspects of the ACC system can be combined and improved by introducing communication with only the preceding vehicle. 
%So, at first the authors focus on demonstrating that the autonomous control law of a typical ACC system cannot ensure an optimal balance between comfort and spacing accuracy. 
%%PRIMA REFERENCE
More recently, the idea of enhancing the control logic of a ACC system by adding communication with only the nearest preceding vehicle was employed also by Naus et al.\cite{naus2010string}. Their control design has been widely used in literature, for example by \cite{lidstrom2012modular,oncu2014cooperative,milanes2013cooperative}. 
They therefore propose a decentralized controller design in which the communication with the nearest preceding vehicle makes the acceleration of the aforementioned vehicle available to the controlled one.
This means that a delay has to be added which in their work is represented by a constant $\theta$.
   
It is interesting to note that, in this case, communication will be implemented as a feedforward signal. Hence, the acceleration of the preceding vehicle is used as a feedforward control signal through a feedforward filter.
A feedforward filter design is thus proposed, based on a zero - spacing error condition: this condition is equivalent to say that the inter-vehicle distance coincides with the desired distance and so the vehicles keep the correct spacing between them.

\vspace{10pt}

The previously described control logic is also present in \cite{naus2009towards}, where it is applied to heterogeneous traffic, focusing in particular on heavy duty vehicles. 
The authors claim that a significant reduction of drag force and so decreasing fuel consumption is expected, but results show that only marginal string stability can be obtained. Therefore, the design cannot be considered robust for uncertainties or errors in the modelling process
Furthermore, Van Der Werf et al.\cite{vanderwerf2001modeling} focus as well on providing the simplest possible cooperative architecture and the easiest one to be implemented.

Their work is then used as base model to develop a CACC control logic by Van Harem et al.\cite{van2006impact} in which a focus on the impact of CACC on traffic - flow characteristics is also present.

The authors finally claim that the CACC system shows potential positive effects on traffic throughput but they also state that the type of communication chosen might be not useful enough to guarantee better traffic-flow performance. 

\vspace{15pt}

Hence, it is clear how results are affected by the type of communication chosen and how different and more complicated information flow topologies should be investigated. 
It is also evident that an innovative and reliable type of communication such as 5G can be beneficial in this case. 
As a consequence of this, Lidstrom at al.\cite{lidstrom2012modular} proposed a modified CACC controller (based on the afore-mentioned work of Naus et al.\cite{naus2009towards} and \cite{naus2010string}) in particular considering the design of the feed forward filter.

\begin{figure}
    \centering
    \includegraphics[scale=0.2]{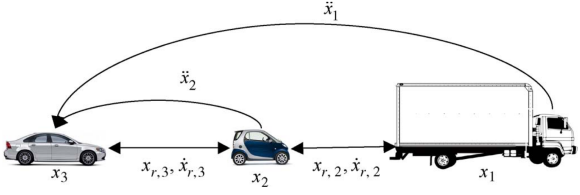}          
    \caption[Vehicles platoon equipped with CACC functionality]{Vehicles platoon equipped with CACC functionality \cite{lidstrom2012modular}}
    \label{vehicles platoon lidstrom}
\end{figure}

As it can been seen in Figure \ref{vehicles platoon lidstrom}, this control strategy differs from the ones of \cite{naus2009towards} for the type of communication: information are taken not only by the directly previous vehicle, but the acceleration signal is transmitted also by the leading vehicle. 

So, essentially:
\begin{itemize}
    \item The feedforward filter is now chosen to achieve string stability, but no conditions on having zero spacing error are stated;
    \item The acceleration to be given in feedforward is chosen to be the minimum between the acceleration of the leading vehicle and the acceleration of the preceding vehicle.
\end{itemize}

Results show that the controller is able to achieve string stability with a certain robustness against dynamics variations and is also able to improve safety during deceleration.

\section{Developed Systems Description}

 \subsection{Commercial Adaptive Cruise Control System}
 \begin{figure}
     \centering
     \includegraphics[scale=0.25]{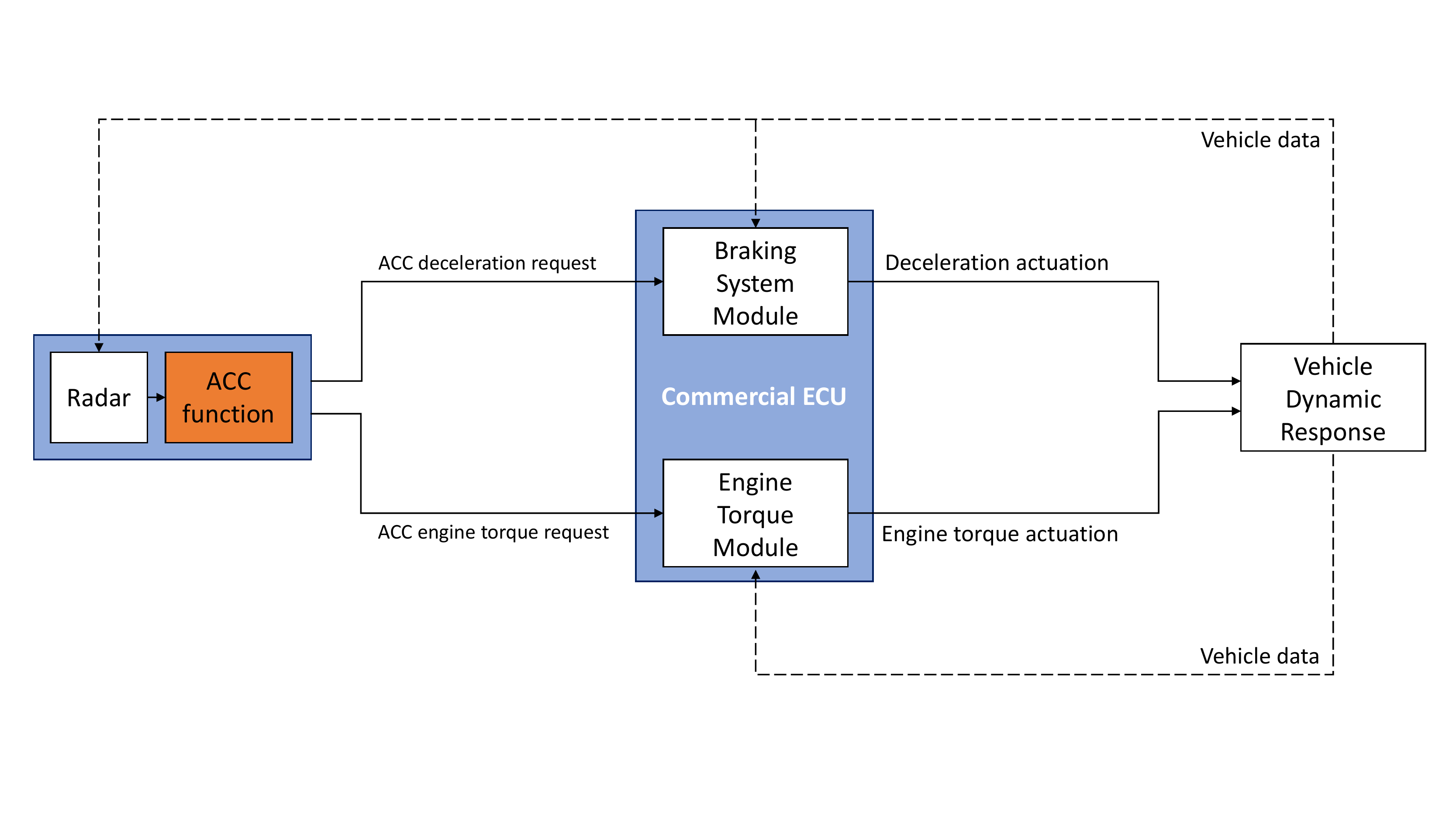}
     \caption{ACC System architecture of a commercial vehicle}
     \label{ACC_FC}
 \end{figure}
The functional architecture of a base case, commercial ACC is shown in Figure \ref{ACC_FC}. It is based on a radar installed on the front part of the vehicle that monitors the road ahead and detect other vehicles. As long as no obstacle is detected, the ACC follows the user-set speed working as a classic Cruise Control in Velocity control. If the system detects a slower object within the radar sensor range, it reduces the speed comfortably either by releasing the gas pedal or by activating the Braking System Module. If the vehicle in front accelerates or change its driving lane, ACC automatically increases its speed up to the user-set speed through the Engine Torque Module. These two tasks are accomplished by a commercial ECU (represented in Figure \ref{ACC_FC}).
Moreover, to perform the overall functioning of the system (and to close the simulation loop as well), a vehicle model is considered, which is represented by the "Vehicle Dynamic Response" block in Figure \ref{ACC_FC}.
 
Afterwards, the commercial ACC system is chosen to be characterized by a hierarchical longitudinal control system architecture, composed by an upper and a lower level controller: 
\begin{itemize}
    \item The upper level controller is the designed ACC controller and determines the desired acceleration;
    \item The lower level controller determines the gas and/or brake commands required to track the provided desired acceleration.
\end{itemize}

\subsubsection{Upper Level Control Logic}
The upper level controller has to provide a desired acceleration command to the lower level control logic. This command is limited between fixed values $a_{Ego,min}^{des}$ and $a_{Ego,max}^{des}$ in order to avoid abrupt manoeuvres. Their values are:
%\begin{frame}
%\footnotesize
\begin{equation*}
    \begin{cases}
    a_{min}^{des}= -5m/s^2\\
    a_{max}^{des}= 2m/s^2    
    \end{cases}
\end{equation*}
%\end{frame}
The maximum boundaries for acceleration are chosen considering the most common values shown in \ref{ACC_sys_arch}. In particular, the values selected by \cite{rajamani2006vehicles,wang2002adaptive,wang2004should} have been chosen.
As already mentioned, two types of controller have to be implemented in order to allow the system to work both in Speed and Spacing Control. 
Speed control is active when the Preceding vehicle is far away from the Ego vehicle or the Preceding vehicle is travelling faster then the Ego car. The control logic is a PI-controller \cite{rajamani2006vehicles} that gets as input the actual car speed and the driver-set speed $v_{user}$. By definition, a PI-controller has no steady state error. Moreover, settling time and overshoot were adjusted by tuning the controller gains.\\
Hence, the desired acceleration in speed control is given by \ref{FCA_CClaw}:
\begin{equation}
    a_{Ego,VC}^{des}(t)=k_{p}(v_{user}(t)-v_{Ego}(t))+k_{i}\int\limits_{0}^{t}(v_{user}(t)-v_{Ego}(t))dt
    \label{FCA_CClaw}
\end{equation}
where $k_{p}$ and $k_{i}$ are results of gain tuning.

\begin{figure}
    \centering
    \includegraphics[scale=0.25]{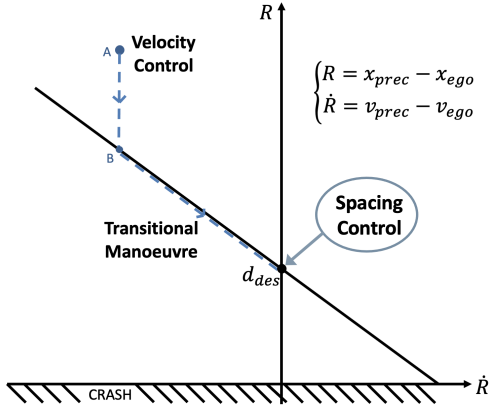}
    \caption{A range versus range-rate diagram: example of linear trajectory along the $R-\dot{R}$ diagram \cite{fancher1994evaluating}}
    \label{fancher_vuoto}
\end{figure}

Before introducing the choice of the spacing controller, it is now important to present the transitional controller design, which is needed in order to switch from Speed Control to Spacing Control. Hence, a range versus range-rate relationship \cite{fancher1994evaluating} has being adopted.
A generic representation of a range versus range-rate diagram is reported in Figure \ref{fancher_vuoto}. 

The conventions used to express range $R$ and range rate $\dot{R}$ are:
\begin{equation*}
    R = x_{Preceding} - x_{Ego}
\end{equation*}
\begin{equation*}
    \dot{R} = v_{Preceding} - v_{Ego}
\end{equation*}
An interesting property of this representation is that on the left half-plane, since $\dot{R}$ is negative, $R$ can only decrease, while on the left half-plane, since $\dot{R}$ is positive, $R$ can only increase.\\ The desired operating condition is represented by point $(0,RH)$. In this point the range $R$ corresponds to the desired distance given by the spacing policy, while the relative speed $\dot{R}$ among the vehicles is zero. 
The reasons why this control strategy %described in %\ref{FCA_trans_man} 
has been selected are the following:
\begin{itemize}
    \item It takes into consideration transitional manoeuvres, so that the desired distance from the Preceding vehicle can be established and maintained smoothly;
    \item Engine and especially brake actuators can saturate, so a solution that avoids hard braking is preferred; 
    \item When a new Preceding vehicle appears, it is not considered as a target vehicle if it is far enough or its speed is higher then the controlled vehicle speed;
    %\item When a new Preceding vehicle appears, it can happen that for its conditions it does not have to be considered as a target vehicle if it is far enough or its speed is higher then the controlled vehicle speed;
    %\item String stability can be enforced if an adequate spacing policy is selected (such as the CTG policy).
\end{itemize}

There are two kinds of transitional manoeuvres according to the considered traffic scenario \cite{fancher1994evaluating}:
\begin{itemize}

    \item When a slower vehicle is encountered, the controlled vehicle performs a linear trajectory in the $R-\dot{R}$ diagram until the steady state condition $RH$ is reached, as shown in \ref{fancher_vuoto}.

    Starting from velocity control, the transitional manoeuvre sub-logic activates when, for a certain $\dot{R}$, the difference between the actual inter-vehicle distance and the corresponding inter-vehicle distance $R_{line}$ on the switching line for the same $\dot{R}$ is smaller than a tolerance, as shown in \ref{FCA_toll_sp_tr}:
    \begin{equation}
        R-R_{line} < tol_{speed-linear}
        \label{FCA_toll_sp_tr}
    \end{equation}
    where it is assumed $tol_{speed-linear}=0.5m$ as a reasonable value during the selection of parameters.
    The term $R_{line}$ refers to the expression of the linear trajectory
    and it is reported in \ref{switchingline_formula}:
    \begin{equation}
        R_{line}=-T\dot{R}+RH
        \label{switchingline_formula}
    \end{equation}
    The slope of the switching line is indicated as $T$ and it is given by \ref{FCA_Tslope} \cite{fancher1994evaluating}:
    \begin{equation}
        T=\sqrt{\frac{R_{s}-RH}{2D}}
        \label{FCA_Tslope}
    \end{equation}
    where:
    \begin{itemize}
        \item $R_{s}$ is the maximum range to consider, in this case the maximum radar sensor range in $[m]$;
        \item $RH$ is the desired inter-vehicle distance in $[m]$ and in the case of constant time gap it depends linearly on the controlled vehicle speed, as shown in \ref{rtkrtk}:
        \begin{equation}
            RH = d_{min}+hv_{Ego}
            \label{rtkrtk}
        \end{equation}
        where $d_{min}=2m$ \cite{rajamani2006vehicles} is the inter-vehicle distance at rest.
        \item $D$ is the coasting deceleration equal to 0.04$g$ $m/s^2$  \cite{fancher1994evaluating}.
    \end{itemize}
    
    A better description of the behaviour of this transitional manoeuvre is possible considering  \ref{fancher_vuoto}. Starting from an initial condition A (where the inter-vehicle distance is high) and considering that the Ego vehicle is travelling in velocity control with a constant negative relative speed with respect to the Preceding vehicle, the sub-logic activates when the trajectory crosses the switching line in point B. The controller then brings the operating condition to spacing control (thus, to point $(0,RH)$) following the slope.\\
    With a linear transitional manoeuvre, the acceleration command is the result of a P-controller and it is shown in \ref{FCA_line}:
    \begin{equation}
        a_{Ego,TC}^{des}(t)=k_{p}(R_{line}-R)
        \label{FCA_line}
    \end{equation}
    where $k_{p}$ is a controller gain. %where $k_{p}=0.03$.
    During the process of controller tuning, it is possible to make the vehicle perform a straight line from B to $RH$  even if some small oscillation is inevitable due to the controller overshoot.
    
    \item When a Preceding vehicle cuts-in with a lower speed then the Ego car, it could be necessary to perform a hard braking. Assuming that the maximum deceleration $a_{Ego,min}^{des}$ of the system is required, this implies a constant deceleration manoeuvre. On the $R-\dot{R}$ diagram this corresponds to a parabolic trajectory as described by \ref{FCA_Rparab}:
    \begin{equation}
        R_{parabola}=R_{amn}+\frac{\dot{R}^2}{2a_{Ego,min}^{des}}
        \label{FCA_Rparab}
    \end{equation}
    where $R_{amn}$ is the minimum of the parabolic trajectory, which occurs when $\dot{R}=0$, as visible in \ref{fancher_parabola}. The condition to activate the parabolic transitional manoeuvre is based on comparing the actual range $R$ with the value of $R$ belonging to the switching line for the same $\dot{R}$. As indicated in \ref{FCA_tol_para}, if their difference is lower then a tolerance, parabolic trajectory sub-logic is activated:
    \begin{equation}
        R-R_{line} < tol_{linear-parabolic}
        \label{FCA_tol_para}
    \end{equation}
    where it is assumed $tol_{linear-parabolic}=-30m$ \cite{fancher1994evaluating}. \\
    Once this transition mode is activated, the acceleration command to be given to the lower level controller is the result of a P-controller (as discussed in \cite{rajamani2006vehicles}). This controller is applied to the relative distance between vehicles and the corresponding $R$ point along the parabola, given a certain $\dot{R}$, as shown in \ref{FCA_ctrl_parab}:
    \begin{equation}
        a_{Ego,TC}^{des}(t)=k_{p}(R_{parabola}-R)
        \label{FCA_ctrl_parab}
    \end{equation}
    where $k_{p}$ is a controller gain. %$k_{p}=0.4$. 
    As said, this parabolic trajectory is likely to occur when a Preceding vehicle cuts-in with a lower speed then the Ego car. Considering \ref{fancher_parabola}, initial condition can be described by point A, where $\dot{R}$ is negative and the desired spacing $R$ is lower then the desired value. The transitional manoeuvre activates according to the condition reported in \ref{FCA_tol_para} and the parabolic trajectory reaches its minimum in correspondence of $R_{amn}$ which has to be positive to avoid collision. In this point the inter-vehicle distance is lower then the desired $RH$. Then the trajectory keeps on following the parabolic path in the right half-plane. Here, a positive $\dot{R}$ means that the Preceding vehicle is driving faster than the controlled one and so this represents a safe situation. In order to reach the final position in point $(0,RH)$, the trajectory continues along the parabola until it crosses the switching line given by \ref{switchingline_formula} in point C. From this point the vehicle performs the linear trajectory until final condition is reached.
    \begin{figure}
        \centering
        \includegraphics[scale=0.32]{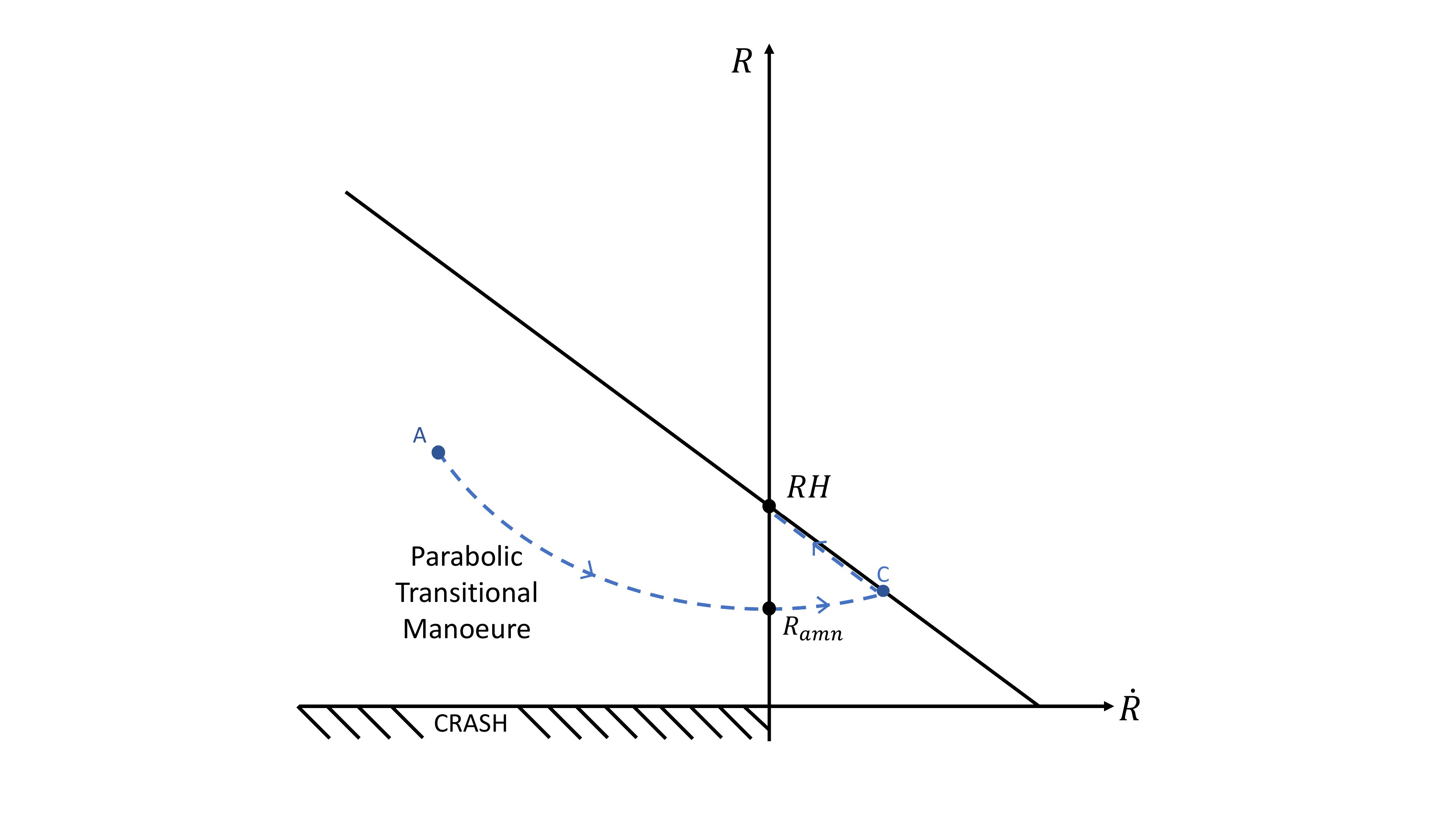}
        \caption{Example of parabolic trajectory along the $R-\dot{R}$ diagram}
        \label{fancher_parabola}
    \end{figure}

\end{itemize}

The sub-logic that contains the spacing control is activated when the operating point in the range versus range-rate diagram is close enough to the desired final condition in point $(0,RH)$:
\begin{equation}
    \vert R-RH \vert < tol_{transitional-spacing}
    \label{}
\end{equation}
where it is assumed $tol_{transitional-spacing}=0.5m$, the same value also given to $tol_{speed-linear}$. The reason for the absolute value in equation is to include both the situations in which during the transitional manoeuvre the operating point can move on the switching line coming from the left half-plane or from the right half-plane. \\ 
The choice of the spacing controller has been made based on a series of considerations. Firstly, it has to satisfy the requirements of string stability and reduction of any distance and speed deviation from reference values. Furthermore, among all the possible control strategies, it is preferred to adopt one that depends on physical quantities, so that it is more intuitive to set the parameters. The selected controller is taken from the work of Winner et al. \cite{winner2016handbook} which has deeply influenced the development of commercial ACC systems \cite{prestl2000bmw}. The controller is represented in \ref{eq_winner}:
\begin{equation}
    a_{Ego,SC}^{des}(t)=\frac{\dot{R}-\frac{d_{set}-R}{\tau_{d}}}{\tau_{v}}=\frac{\dot{R}}{\tau_{v}}-\frac{d_{set}-R}{\tau_{v}\tau_{d}}
    \label{eq_winner}
\end{equation}
where $\tau_{d}$ and $\tau_{v}$ are parameters and $d_{set}$ is defined according to the constant time gap spacing policy, as shown in \ref{eq_CTH_FCA}:
\begin{equation}
    d_{set}=d_{min}+hv_{Ego}
    \label{eq_CTH_FCA}
\end{equation}
where $d_{min}=2m$ \cite{rajamani2006vehicles} is the inter-vehicle distance at rest. In order to guarantee string stability, the condition on the parameters in \ref{tau_string_stab} has to be satisfied according to \cite{winner2016handbook}:
\begin{equation}
    \tau_{v} \leq \tau_{gap} \left( 1+\frac{\tau_{gap}}{2\tau_{d}} \right)
    \label{tau_string_stab}
\end{equation}
As described in \cite{winner2016handbook}, the control approach represented by \ref{eq_winner} is string stable and also meets the requirements of a constant time gap only if the condition in \ref{tautau} is satisfied:
\begin{equation}
    \tau_{v}=\tau_{set}
    \label{tautau}
\end{equation}
Therefore, keeping into account \ref{tau_string_stab}, the condition on $\tau_{d}$ in order to have string stability is shown in \ref{taud}:
\begin{equation}
    \tau_{d} \geq 0
    \label{taud}
\end{equation}
The value for $\tau_{d}$ is selected considering the scenario described in \cite{winner2016handbook}: a vehicle cuts in without a speed difference at a distance that is $\gamma=20m$ smaller the the desired distance. Considering a reasonable deceleration of $1m/s^2$ to be applied to the vehicle, which corresponds to take the foot off from the gas pedal or slightly braking, according to \ref{eq_winner} it is possible to find a value for $\tau_{d}$ inversely proportional to $\tau_{v}$, as seen in\ref{taudgamma}:
\begin{equation}
    \tau_{d}\tau_{v}=\gamma \implies \tau_{d}=\frac{\gamma}{\tau_{v}}
    \label{taudgamma}
\end{equation}

\subsection{CACC system}
\begin{figure}
    \centering
    \includegraphics[scale=0.25]{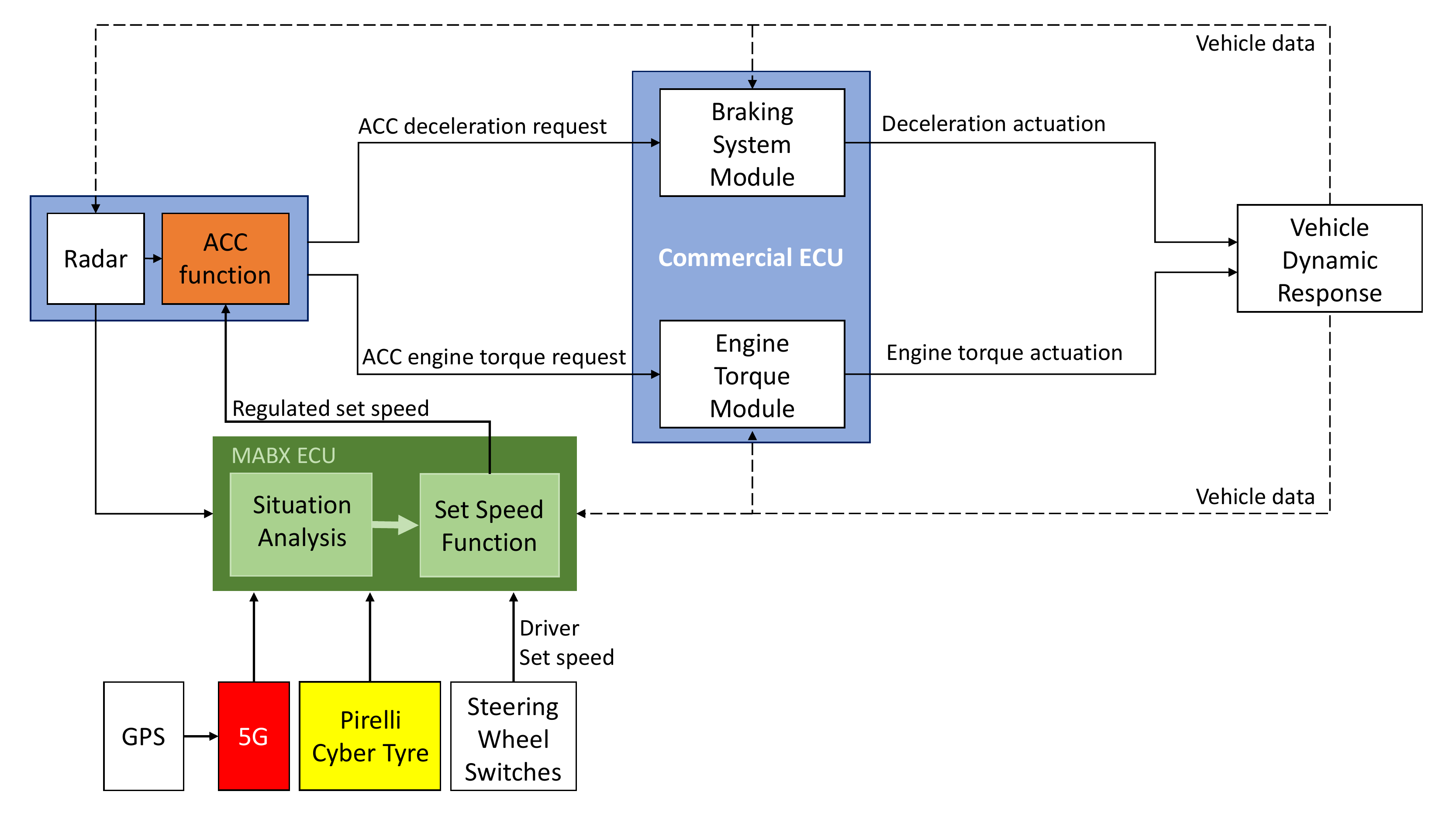}
    \caption{Connected ACC system architecture}
    \label{CACCfirst}
\end{figure}

The architecture of the novel CACC system is shown in \ref{CACCfirst}. 
The approach adopted is to consider the base case commercial ACC logic previously described as the main logic on which the novel developed functionalities are connected as Add-Ons. These additional features are able to extend the potentialities of the original ACC system. 
Hence, the original ACC function block is considered as a fixed component and consequently is not modified, but the logic received now information coming from the 5G communication block and from the Cyber Tyre block. 
This approach is motivated by the following considerations: \begin{itemize}
    \item The implementation of the Connected ACC is much faster and not very demanding in term of cost and time for design since the original ECU of the vehicle does not need to be redesigned or substituted. Therefore, this solution is ready to be introduced in the market in a short time;
    \item The original commercial ACC system is a logic that has been already tested and has proved to work correctly in a wide range of scenarios.   
\end{itemize}

\begin{figure}
    \centering
    \includegraphics[scale=0.25]{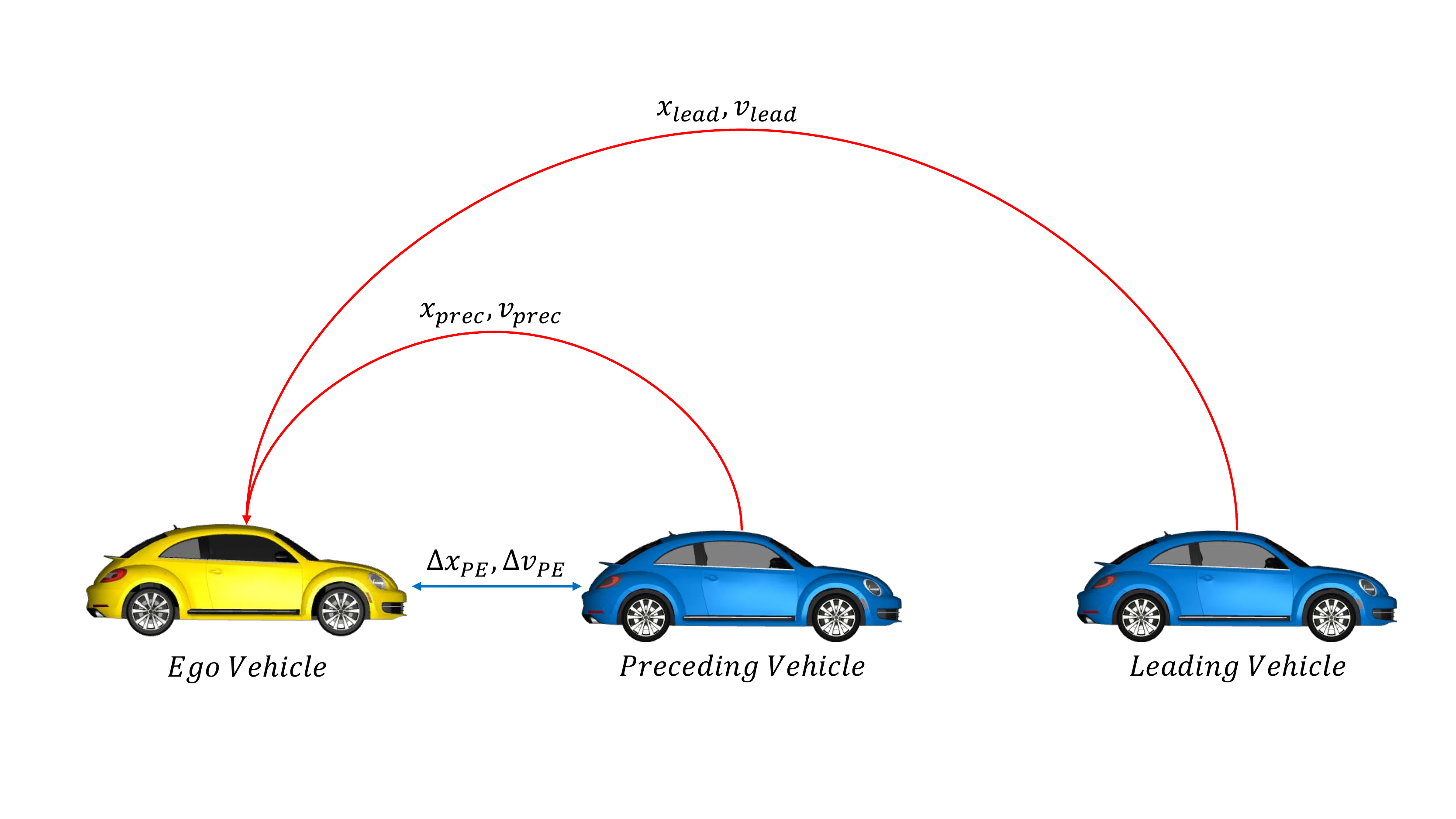}
    \caption{Information Flow Topology}
    \label{3vehicleflowinfoFCA}
\end{figure}

Considering a string of three vehicles as shown in \ref{3vehicleflowinfoFCA}, it is assumed that real time positioning, speed and acceleration of the Preceding and Leading vehicles through GPS information can be transmitted through the 5G network. Hence, the Ego vehicle receives information both through 5G communication (displayed with the red arrows) and the radar system (displayed with the blue arrow). 

At this point, from the exchanged data between the Ego and the Leading vehicle, a a Time-To-Collision ($TTC$) has been built. The use of $TTC$ as safety indicator is widespread in literature. It was firstly introduced by Hayward \cite{hayward1972near} and later developed by Hyden \cite{hyden1987development}, founding a great success in collision avoidance problems such as \cite{vogel2003comparison,van1993time,minderhoud2001extended} because of its ability to consider the distance between vehicles weighted on their relative speed. Therefore, a risky situation occurs when $TTC$ is too low which is the result either of a much bigger $v_{Ego}$ compared to $v_{lead}$ or of a too small inter-vehicle distance. A threshold value $TTC_{lim}$ is taken as indicator for the severity of the scenario and its value ranges from $1.5s$ to $4s$\cite{minderhoud2001extended}. A high $TTC_{lim}$ can result in generating too many false alarms while, in the case of autonomous vehicles, lower values are allowed due to a more adequate response in case of dangerous situations \cite{minderhoud2001extended}.\\
In case the TTC is computed between two vehicles separated by another vehicle, not only two inter-vehicle spacing have to be considered, but also the length of the car in-between. Therefore, it is suggested to assume $TTC_{lim}=6s$. 

In this work, the inverse of the Time-To-Collision $TTC^{-1}$ is considered. Its definition has the same physical meaning of the $TTC$, but the expression is different. The inverse of the time to collision of the controlled vehicle with respect to the Leading vehicle is represented in \ref{TTCEL}:
\begin{equation}
    TTC_{E,L}^{-1}=\frac{\Delta v_{E,L}}{\Delta x_{E,L}}
    \label{TTCEL}
\end{equation}
where:
%\begin{numcases}{}
\begin{equation}
   \Delta v_{E,L} = v_{Ego}-v_{lead}\\
\end{equation}
\begin{equation}
   \Delta x_{E,L} = x_{lead}-x_{Ego}
\end{equation}
%\end{numcases}
The threshold value to be considered therefore is:
\begin{equation}
    TTC_{lim}^{-1}=\frac{1}{TTC_{lim}}=\frac{1}{6}s
\end{equation}

\begin{figure}
    \centering
    \includegraphics[scale=0.2]{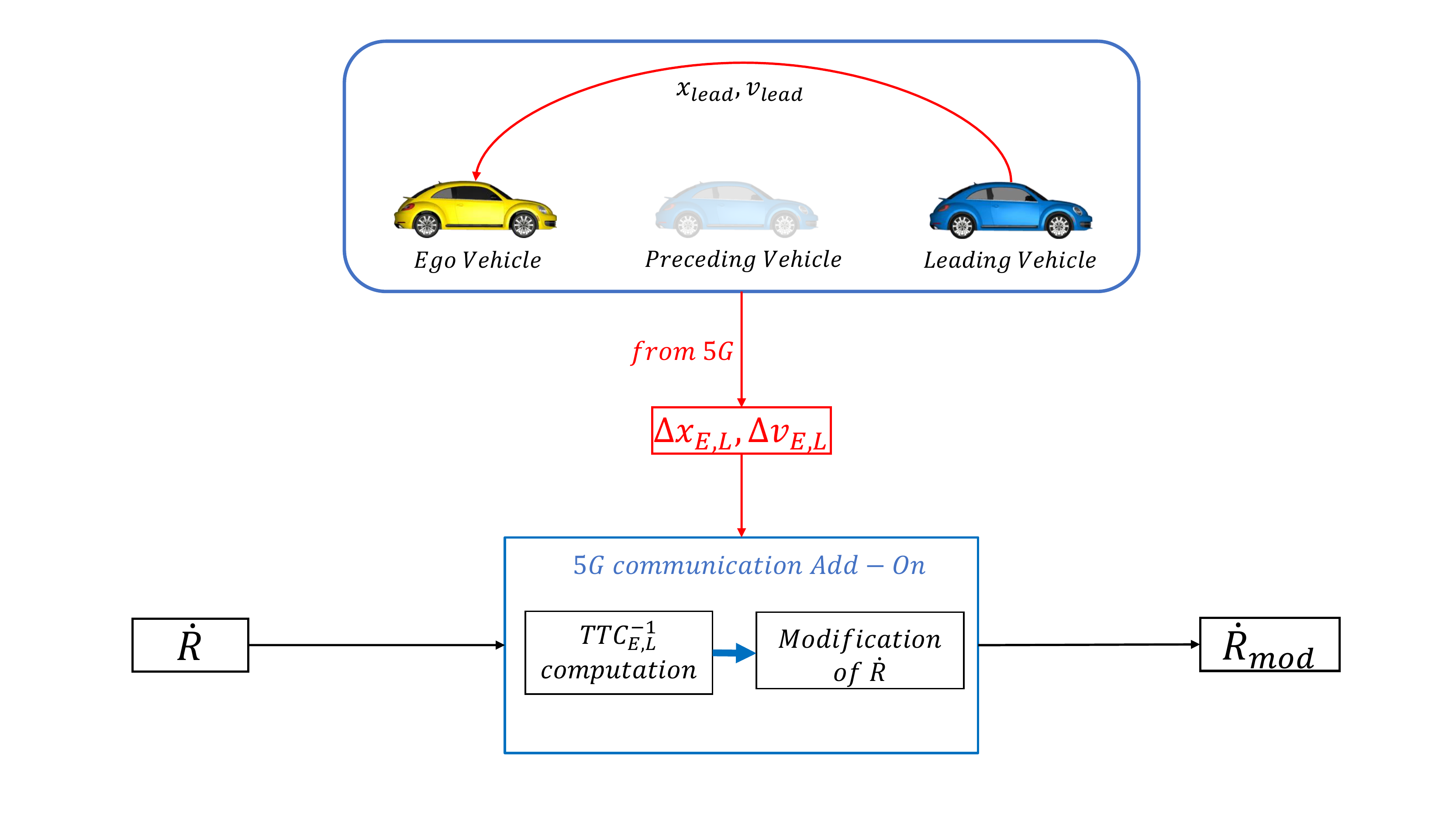}
    \caption{Representation of the modification to $\dot{R}$}
    \label{V2V}
\end{figure}

The information coming from 5G network and the computation of the $TTC^{-1}$ are used to modify the relative speed between the Ego vehicle and the Preceding one when the cars are in Spacing Control. In \ref{V2V} the representation of the modification to $\dot{R}$ is showed.
\vspace{1pt}

In the case where that $TTC_{E,L}^{-1}$ is lower then the threshold value, the modifications to the relative speed between Ego and Preceding vehicle is given by \ref{FCA_R_mod}:
\begin{equation}
    \dot{R}_{mod}=\left( 1-sgn(\dot{R})\frac{\Delta v_{E,L}}{n_{1}} \right) \dot{R}
    \label{FCA_R_mod}
\end{equation}
where $n_{1}$ is a normalization factor and it is represented by the imposed speed limit of the road. 
\begin{figure}
    \centering
    \includegraphics[scale=0.25]{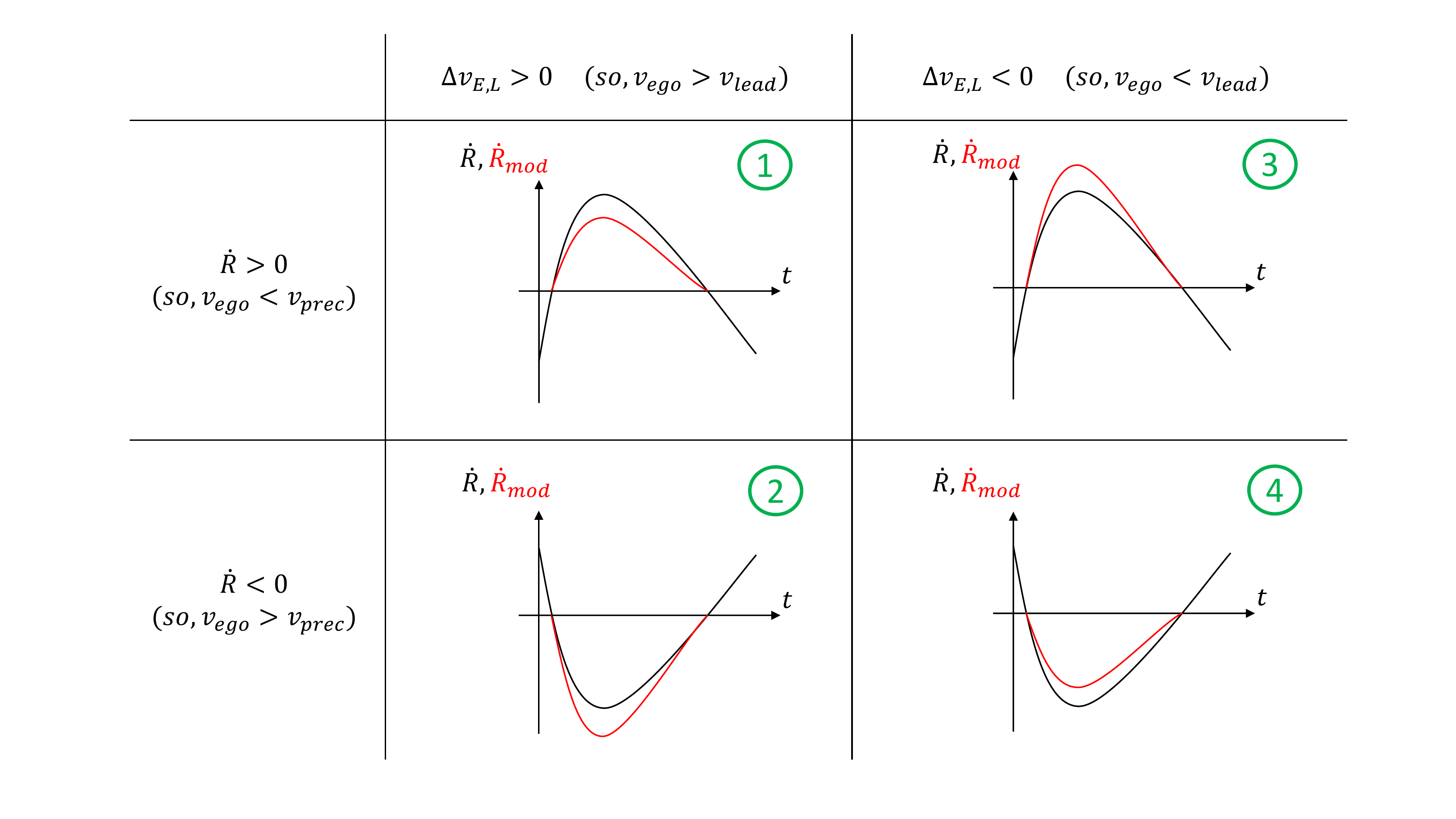}
    \caption{Effect of $\dot{R}$ modification according to different situations}
    \label{tableRmod}
\end{figure}

In \ref{tableRmod}, the physical meaning of the modification to $\dot{R}$ for the Ego vehicle is explained. In particular, it can be stated that the control logic is able to perform in a wide range of scenarios:
\begin{itemize}
        \item In case of a positive $\Delta v_{E,L}$ (therefore $v_{Ego}>v_{lead}$), the Leading vehicle is travelling slower then the Ego vehicle. Therefore, a preventive action is desired in order to avoid possible dangerous scenarios. If the speed of the Preceding vehicle is higher than that of the Ego vehicle (situation \textcircled{1}),
        a potential variation of $\dot{R}$ is smoothed. In fact, since the Leading vehicle is slower, it is predictable that soon also the Preceding vehicle is going to slow down. This modification to $\dot{R}$ can allow to avoid unnecessary acceleration command.
        \vskip 2pt
        On the other hand, if the Ego car is faster than the Preceding one (situation \textcircled{2})
        , the modified $\dot{R}_{mod}$ is higher in absolute value with respect to the original $\dot{R}$. Considering a scenario where the Leading vehicle starts decelerating, the advantage of adopting this solution is that a connected Ego vehicle is able to anticipate the braking manoeuvre, so that it does not have to wait for the car in-between to react. This can prevent risk of collision.%\\
        
        \item In case of a negative $\Delta v_{E,L}$ (therefore $v_{Ego}<v_{lead}$), the Leading vehicle is increasing its distance from the Ego vehicle. If the Preceding vehicle is faster than the Ego (situation \textcircled{3})
        , the modified $\dot{R}_{mod}$ is higher than $\dot{R}$ so that a higher acceleration is produced as output by the control logic. This can result in a vehicles being closer and therefore a higher traffic capacity.  
        \vskip 2pt
        On the other way, when the Preceding vehicle travels at a lower speed then the Ego (situation \textcircled{4})
        , the value for $\dot{R}$ is decreased in absolute value so that the control logic can provide a smoother deceleration. This does not represent a danger in case of hard-braking since in \ref{FCA_R_mod} the behaviour of the Preceding vehicle still plays a major role.
        
        \item In the case that the Leading and Ego vehicle are travelling at the same speed, so $\Delta v_{E,L}=0$, the input to the vehicle-following control logic does not change and therefore the inter-vehicle distance matches with the distance given by the constant time gap policy.

\end{itemize}
\vspace{1pt}
In case that $TTC_{E,L}^{-1}$ is higher then the threshold value, the modifications to the relative speed between Ego and Preceding vehicle is given by the following equation \ref{FCA_R_mod_ttc}:
\begin{equation}
    \dot{R}_{mod}=\left( 1-sgn(\dot{R})\frac{\Delta v_{E,L}}{n_{1}}+\frac{TTC_{E,L}^{-1}-TTC_{lim}^{-1}}{n_{2}} \right) \dot{R}
    \label{FCA_R_mod_ttc}
\end{equation}
where $n_{1}$ and $n_{2}$ are normalization factors. As concerns $n_{1}$, it holds what explained for \ref{FCA_R_mod}.
The value assigned to $n_{2}$ is assumed to be $TTC_{lim}^{-1}$, so that all the multipliers of $\dot{R}$ in \ref{FCA_R_mod_ttc} are comparable in term of order of magnitude. It is important to underline that in the case where \ref{FCA_R_mod_ttc} is applied, $\Delta v_{E,L}$ is always positive. Comparing this formulation to \ref{FCA_R_mod}, 
an additional term that considers the difference between the actual $TTC_{E,L}^{-1}$ and the threshold value is introduced. Since in this case this difference is always positive, the higher the difference, the higher $\dot{R}_{mod}$. So, not only a smoother deceleration is guaranteed, but it is also possible to mitigate the risk of collision. 

When the controlled vehicle has only one vehicle ahead, the value for $TTC_{E,L}^{-1}$ does not influence the controller. In this case, the data transmitted to the 5G network by the Preceding car are compared to the one measured by the radar sensor. In particular, the relative speed given as input to the controller is the highest in absolute value comparing the one observed by the radar and the one received through V2V communication, so that the worst case scenario is considered by the algorithm.

In addition to this, it is important to notice that the potential grip is still unknown in initial conditions because state-of-the-art Cyber Tyres can provide the value of the estimated potential adherence during a braking manoeuvre. At the beginning, the initial value of potential grip is assumed as a reference value (like in commercial systems), while it is updated by the estimated value by the system after the deceleration reached a certain threshold.
%se necessario mettiamo figura di come aggiorna il valore
In this work it is supposed that Cyber Tyre system has already calculated the potential grip in a preceding braking manoeuvre, so that during the simulation the value for asphalt adherence is already estimated.

The value for potential grip is used to change the limit values $a_{ego,min}^{des}$ and $a_{ego,max}^{des}$ for the desired acceleration or deceleration. Their values are:
\begin{equation*}
    \begin{cases}
    a_{min}^{des}= max(-5;-\mu g) m/s^2\\
    a_{max}^{des}= min(2;\mu g) m/s^2    
    \end{cases}
\end{equation*}
This is done in order not to overcome the threshold related to the maximum allowed longitudinal acceleration and deceleration. The vehicle can thus avoid possible tyres forces saturation.

\begin{figure}
    \centering
    \includegraphics[scale=0.22]{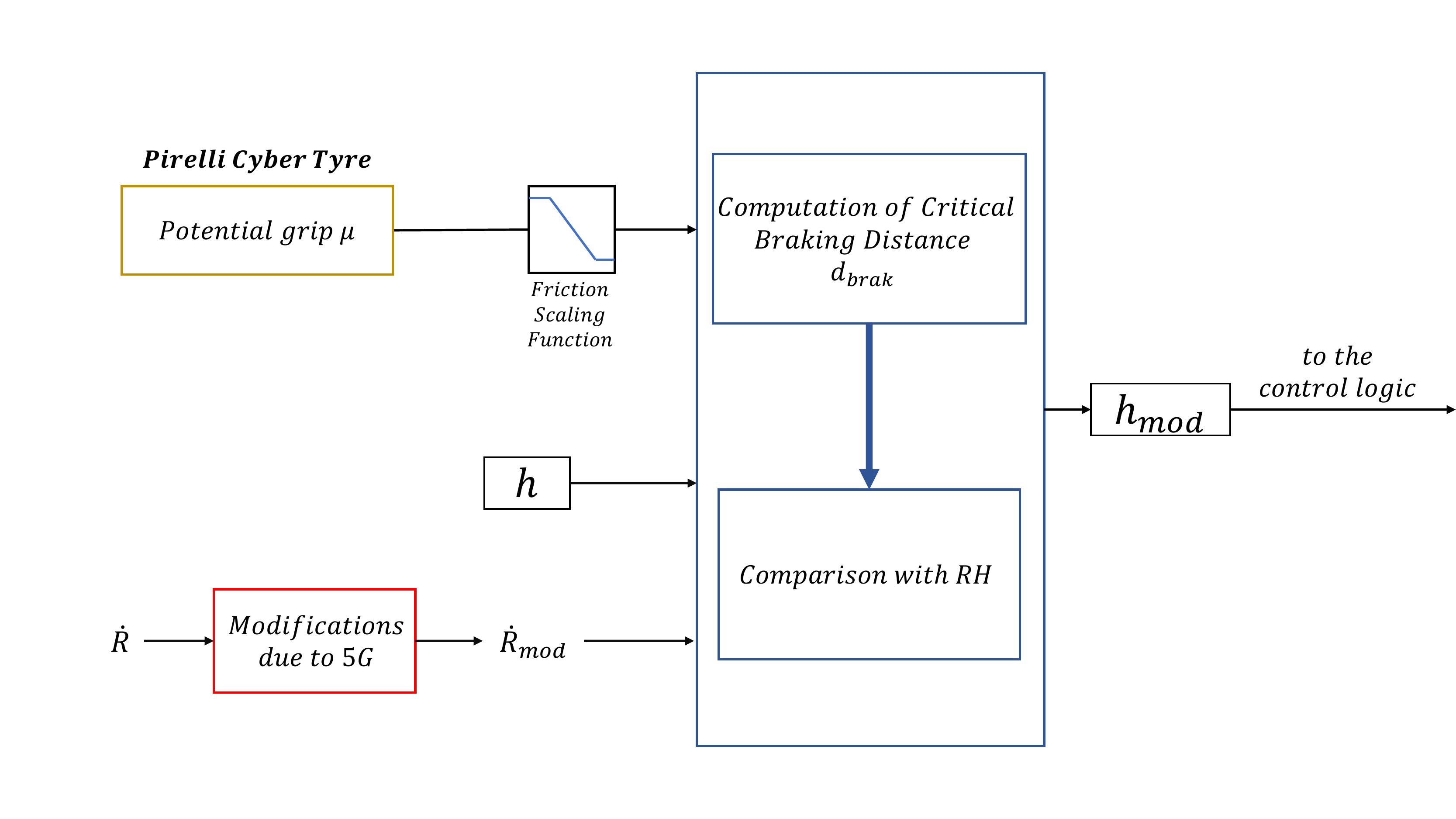}{\centering}
    \caption{Representation of the modification to time gap $h$}
    \label{CT}
\end{figure}

Moreover, the estimated value for grip is used to modify the time gap that the controlled vehicle uses to regulate its distance from the preceding one. In fact, in case of wet or icy road, the braking capabilities of the system cannot be enough to avoid a collision, so that the inter-vehicle distance must be increased. Therefore, the reference $RH$ is compared with a braking critical distance $d_{brak}$ \cite{moon2009design}, expressed by \ref{dbrak_FCA}:
\begin{equation}
    d_{brak}= f(\mu)\left( \frac{v_{ego}^{2}-( v_{ego}+\dot{R}_{mod} )^{2}}{2\vert a_{ego,min}^{des}\vert} \right)-\tau_{s,delay}\dot{R}_{mod}
    \label{dbrak_FCA}
\end{equation}
In this expression, $\tau_{s,delay}$ is the estimated braking actuators delay, $\dot{R}_{mod}$ is the relative speed with the modifications due to the V2V communication %discussed  \ref{5g_fca}
and $f(\mu)$ is a piece-wise linear function used for the friction scaling \cite{yi1999study}. The latter is is shown in %\ref{mu_frict}
and it is described as:
\begin{equation} 
    \begin{cases}{f(\mu)=}
        f(\mu_{min}) & if \mu \leq \mu_{min}  \\ 
            f(\mu_{min})+\frac{f(\mu_{norm})-f(\mu_{min})}{\mu_{norm}-\mu_{min}}(\mu-\mu_{min}) & if \mu_{min} < \mu < \mu_{norm} \hspace{0.9cm} \\ 
            f(\mu_{norm}) & if \mu \geq \mu_{norm}  
    \end{cases}
\end{equation}
where $\mu_{min}=0.2$ is the smallest friction coefficient to be considered and $\mu_{norm}=0.9$. In particular, $f(\mu_{norm})$ is set to unity and $f(\mu_{min})$ is set to $\frac{\mu_{norm}}{\mu_{min}}=4.5$ \cite{moon2009design}. So, if friction decreases, the scaling factor $f(\mu)$ increases and consequently also $d_{brak}$.
   \begin{figure}
    \centering
        \includegraphics[scale=0.3]{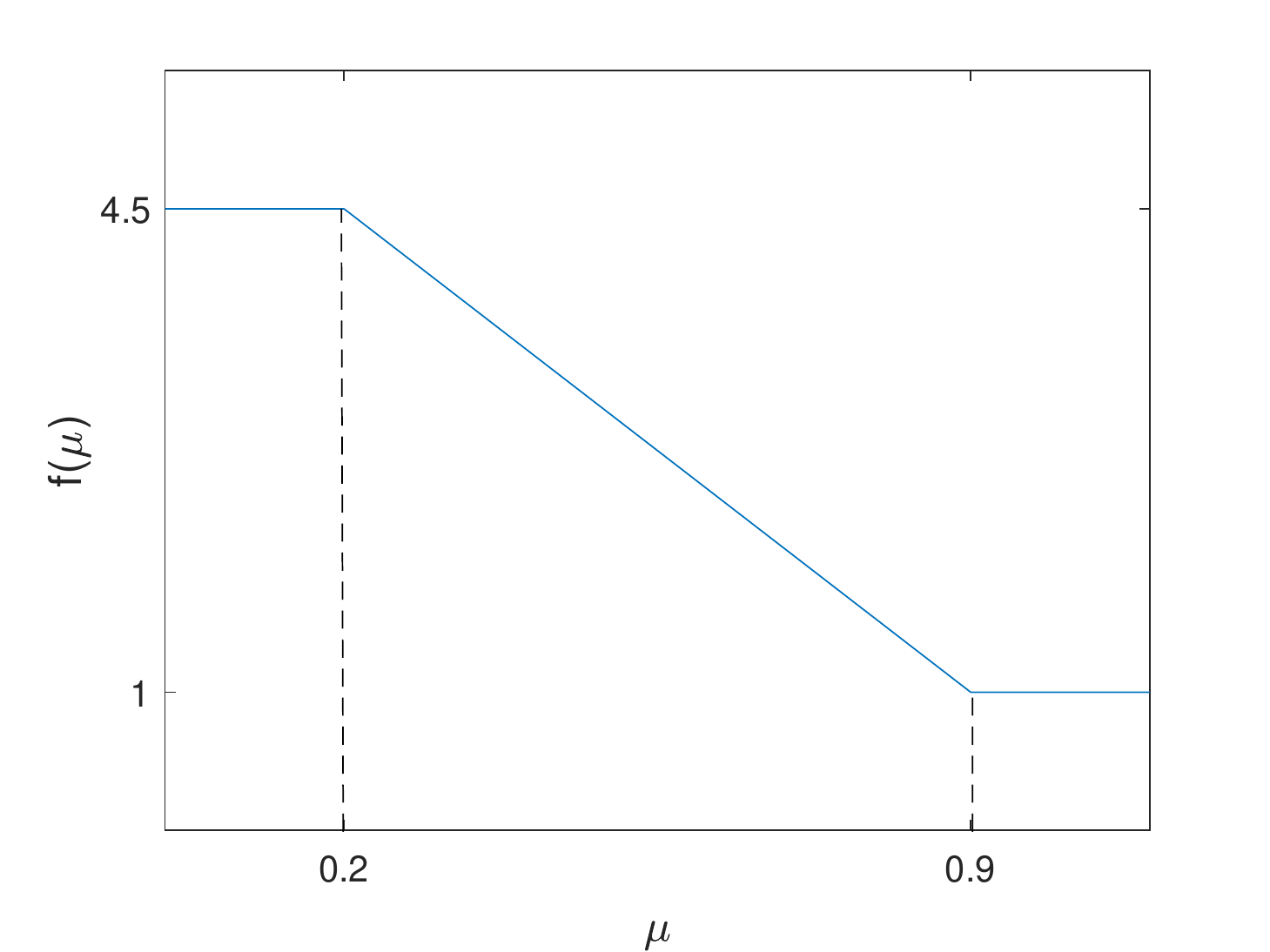}          
        \caption[Friction scaling function]{Friction scaling function}
        \label{mu_frict}
\end{figure}\\
The controller then compares the computed $d_{brak}$ with the $RH$ given by the spacing policy. The modified value for the time gap $h_{mod}$ is:
\begin{numcases}{h_{mod}=}
    h & if $d_{brak} < RH$\\
    \frac{d_{brak}-d_{min}}{v_{ego}} & if $d_{brak} > RH$
\end{numcases}
where $d_{min}=2m$ \cite{rajamani2006vehicles} is the inter-vehicle distance at rest.
The described procedure can be summed up in the representation in \ref{CT}.

Another possible way to take advantage of the information of potential grip is modify the minimum distance in the constant time gap policy $d_{min}$. In particular, in those situations where $d_{brak}>d_{min}$, the minimum distance could be substituted by the braking critical distance so that safety can still be guaranteed. However, such a modification can compromise the benefit of using a Constant Time Gap (CTG) policy, which is string stable. In fact, the term $d_{brak}$ is not linearly dependent on the $v_{ego}$ and it is function also of $\mu$ and $\dot{R}_{mod}$. Therefore, this alternative use of the information of potential grip is only mentioned but it will not be discussed later in this work.

\section{Scenarios Description}
The scenarios considered for systems performance evaluation and comparison are now briefly described. 

Simulations will be divided into a "Vehicle Approaching and Sudden Braking Manoeuvre" and an "Oscillatory Movement of the Leading Vehicle". When dealing with connected systems, the information flow topology chosen refers to Two Predecessors Following Topology.

\subsection{Vehicle Approaching and Sudden Braking Manoeuvre}
\label{breaking_scenario}

This kind of braking manoeuvre is a typical situation on a highway \cite{rajamani2002semi}. The use of this particular stressful scenario is widespread in literature both in the ACC and the CACC field: as an example, Martinez at al. \cite{martinez2007safe} and Nilsson \cite{nilsson1996safety} made high use of this type of scenario in testing the ACC system, while Rajamani et al. \cite{rajamani2002semi} uses the hard braking scenario to test string stability in a semi-autonomous adaptive cruise control system as outlined in \ref{Cap2_CACC_Classic control systems}.
In many other works heavy braking manoeuvre are included in more complex speed/acceleration profiles in order to test the robustness of CACC systems \cite{schmied2015nonlinear,5531155,wang2019cooperative}. 
Starting from these considerations, the developed scenario is characterized by three vehicles:
\begin{itemize}
    \item Leading vehicle with imposed motion;
    \item Preceding vehicle with the selected control logic;
    \item Ego vehicle with the selected control logic.
\end{itemize}
\begin{figure}
    \begin{center}
        \includegraphics[scale=0.25]{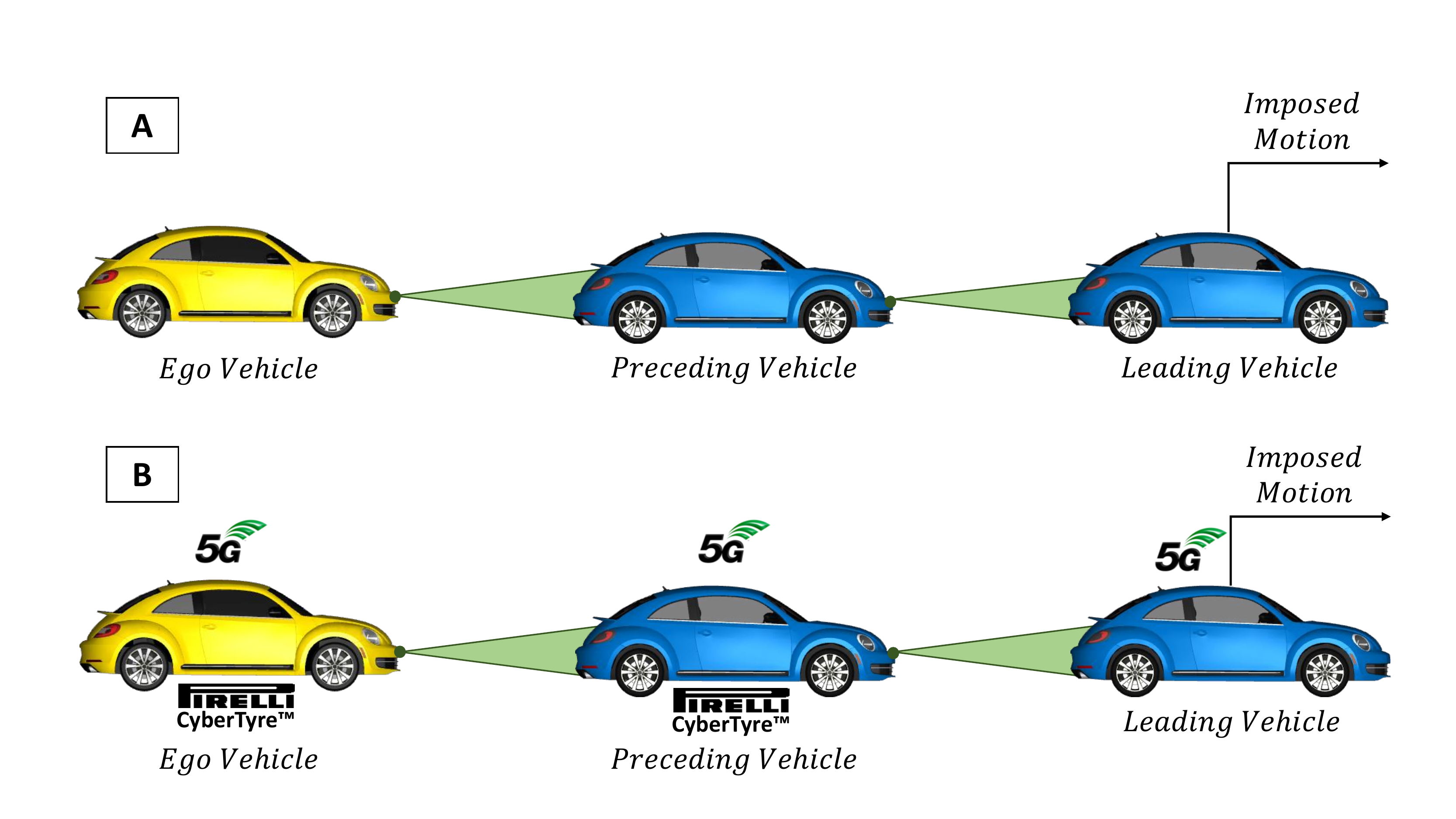}
          \caption{Vehicles set-up}
          \label{3vehicleflowinfoMPC_vuoto}
    \end{center}
\end{figure}

In \ref{3vehicleflowinfoMPC_vuoto} the spacial order of the vehicles is reported: in configuration A only radar is present, while in configuration B also 5G communication and Pirelli Cyber Tyre are available. It is important to specify that both the Ego and the Preceding vehicle are equipped with the same control logic.
%si capisce che intendo sia per ACC che CACC? 

The three vehicles start at a initial distance of $140m$, so that the vehicles can exhibit the behaviour in Cruise Control. The leading vehicle drives at a constant speed of $90 km/h$ while the following vehicles (the Ego and the Preceding vehicle) reach the user set speed of $v_{user}=100km/h$. After the vehicle in front has been detected, the Preceding and Ego vehicles perform a transitional manoeuvre until the inter-vehicle distance with the vehicle in front corresponds to the one established by the spacing policy. Just before $t_{braking}=150s$, all the vehicles are in steady-state following condition. At that exact instant, the leading vehicle performs a sudden braking, from $90km/h$ to $30km/h$ with a deceleration of $-4.5m/s^2$.
It is worth noting that hard braking scenario is also a suitable situation to highlight the limitation of the commercial ACC system as Nilsson \cite{nilsson1996safety} points out in his work. Furthermore, the scenario will be analyzed varying the friction coefficient of the road in order to highlight the action of the Cyber Tyre. It is assumed that for dry surface $\mu_{dry}=0.8$ and for wet surface $\mu_{wet}=0.5$ \cite{cheli2015advanced}. Concerning Cyber Tyre, it is supposed that the potential grip estimation system has already evaluated the asphalt adherence in a preceding braking manoeuvre, so that during the simulation the value for potential grip is already estimated.

\subsection{Oscillatory Movement of the Leading Vehicle}\label{scen_oscil}
The goal of this scenario is the study of string stability of a line of vehicles, together with the evaluation of ride comfort in terms of acceleration amplitude. 
An oscillatory movement (sine wave) is imposed to the leading vehicle in order to mimic traffic oscillations often present on highways. Many authors in their works perform string stability experiments in scenarios with roughly regular oscillations of the leading vehicle in real traffic conditions \cite{naus2010string,milanes2014modeling,oncu2014cooperative}.
According to \cite{zielke2008empirical}, traffic oscillations can be caused by a series of reasons, among them lane-changing and car-following behaviour, and they are characterized by the following parameters:
\begin{itemize}
    \item Amplitude (\textit{Ampl}): Zielke at al. \cite{zielke2008empirical} define the amplitude of a traffic oscillation as one-half the difference between the maximum observed speed and the minimum observed speed: 
    $\frac{1}{2}(v_{max}-v_{min)}$
    
    \item Period (\textit{T}): regular oscillations can be assumed in this scenario as Schonhof et al. \cite{schonhof2007empirical} demonstrate in their work.
\end{itemize}

For the work purpose, an 8 vehicles string is taken into account, as it has been used in literature to evaluate performances of controlled vehicles \cite{rajamani2001experimental, wang2002adaptive}. All the vehicles are equipped with the same control logic.

The characteristics of the oscillatory scenarios are shown in Table  \ref{tableoscill}: 

\begin{table}
\footnotesize
    \centering
    \scriptsize
     \begin{tabular}{c c c c} 
     \hline
     Scenario Name & Velocity Mean Value & Amplitude & Sine Wave Period\\
     \hline
     Oscillatory 1 & $80 km/h$ & $Ampl= \pm4 km/h$ & $T=40s$\\
    
     Oscillatory 2 & $80 km/h$ & $Ampl= \pm4 km/h$ & $T=20s$\\

     \hline
    \end{tabular}
    \caption{Oscillatory scenarios}
    \label{tableoscill}
\end{table}

The \textit{Oscillatory 1} scenario is the base case. The developed system is tested in a more severe situation with respect to the typical traffic scenario, which are characterized by a period around $3-4 min$ \cite{zielke2008empirical}. Finally, the amplitude of velocity oscillations is taken from the work of Zheng et al.\cite{zheng2010impact}.

The \textit{Oscillatory 2} scenario has the same amplitude of the base case but with a halved sine wave period.

The road friction is not a relevant parameter in this kind of scenario, so it is assumed to be in dry surface conditions where $\mu_{dry}=0.8$.

\section{IPG CarMaker/Simulink Implementation and Vehicle Setup}
CarMaker is a simulation tool developed by IPG Automotive, which allows to recreate realistic scenarios in a virtual environment. This software includes a complete model environment comprising an intelligent driver model, a modifiable vehicle model and flexible models for roads and traffic. The software can perform virtual testing of light-duty vehicles with the possibility of recreating every type of road and traffic by means of a Virtual Vehicle Environment (VVE). 
CarMaker provides a user-friendly Simulink interface through the option \textit{CarMaker for Simulink}. This latter is a complete integration of CarMaker vehicle dynamics simulation software into the MathWorks modeling and simulation environment MATLAB/Simulink \cite{quickstartguide}.

\begin{figure}
    \centering
    \includegraphics[scale=0.5]{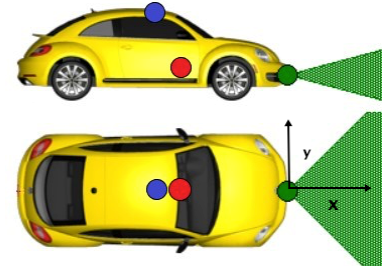}
    \caption[Sensor set-up]{Sensor set-up: the green, the red and the blue dot represent respectively the radar, the GPS and the inertial sensor}
    \label{Sensor_arrangement} 
\end{figure}

The vehicle selected for the scenarios simulations is a 2012 Volkswagen New Beetle, as shown in \ref{Sensor_arrangement}. Furthermore, it is important to consider that:
\begin{itemize}
    \item The vehicle is loaded with an additional weight of $160kg$ (2 passengers, $80kg$ each) in order to simulate a more realistic situation;
    \item The vehicle is provided with an automatic transmission: the ACC (and thus the proposed CACC systems) works in a wide range of velocities, thus requiring gears changing \cite{prestl2000bmw}.
\end{itemize}
The vehicle is equipped with the following sensors:
\begin{itemize}
    \item Radar sensor
    \item Inertial sensor
    \item GPS sensor
    \item Cyber Tyres (only in case of Connected systems)
\end{itemize}
While  Radar, Inertial and GPS sensors are modelled in CarMaker, a model for Cyber Tyres has to be developed.
As concerning 5G communication, the information coming from the preceding and the vehicle in front of the preceding one is transmitted via 5G network. The main parameters considered in this work regarding this communication technology are the following:
\begin{itemize}
    \item Average Latency: $\Bar{l}ms$ 
    \item Minimum Latency: $l_{min}ms$
    \item Maximum Latency: $l_{max}ms$
    \item Latency standard Deviation: $l_{std,dev}ms$
\end{itemize}

As every kind of communication, also 5G suffers from a certain package loss. The estimated percentage of package loss will be addressed as $PL$ due to confidentiality reasons. In order to model this very small error, the following strategy is followed:
\begin{itemize}
    \item  A MATLAB function is created in which a random number \textit{N} is generated;
    \item  The number \textit{N} is compared with a threshold value fixed to $1 - PL$;
    \item If \textit{N} is greater than the threshold value, the delay due to package loss will be added to the Latency delay. Otherwise, the delay of the 5G communication will be made up only by the latency of the signal.
\end{itemize}

\section{Virtual Testing Results}
The Commercial ACC and the Connected ACC systems are virtually tested in order to highlight the advantages of connectivity and the use of smart sensors. 
\subsection{Vehicle Approaching and Sudden Braking Manoeuvres}
The comparisons shown in this section refer to the scenario described in \ref{breaking_scenario}. It is important to note that the commercial ACC system with $h = 0.6 s$ cannot be implemented in a real scenario since it is characterized by string instability, while the connected system have shown to achieve string stability also with $h = 0.6 s$. 

Therefore, a "Commercially available candidate" comparison is carried out in which the control logics adopt the time gap required for string stability requirements. 
Thus, the developed control logics are compared considering systems that can be actually implemented in commercial applications, which results in using a time gap $h = 1.1 s$ for the ACC system and $h = 0.6 s$ for the Connected one. 
For each control logic, the comparison concerns the following profiles:
\begin{itemize}
    \item Acceleration profile of the Ego vehicle;
    \item Velocity profile of the Ego vehicle;
    \item Relative distance profile between Ego and Preceding vehicles;
    \item Spacing error profile between Ego and Preceding vehicles;
\end{itemize}
The purpose of these tests is to show how safely the developed control logics can operate. Therefore, the plots for relative distance and $TTC^{-1}$ are suitable to asses the improved safety features of the connected systems with respect to the non-connected ones. At the same time, concerning the level of comfort during the braking manoeuvre, the comparisons for the acceleration, velocity for the different control logics are provided.

\subsubsection{Normal Ground Friction}

The tests presented were conducted considering a normal ground friction coefficient set to $\mu_{dry} = 0.8$. 

\begin{figure}
    \centering
    %\captionsetup[subfigure]{justification=centering}
    \includegraphics[width=0.8\linewidth]{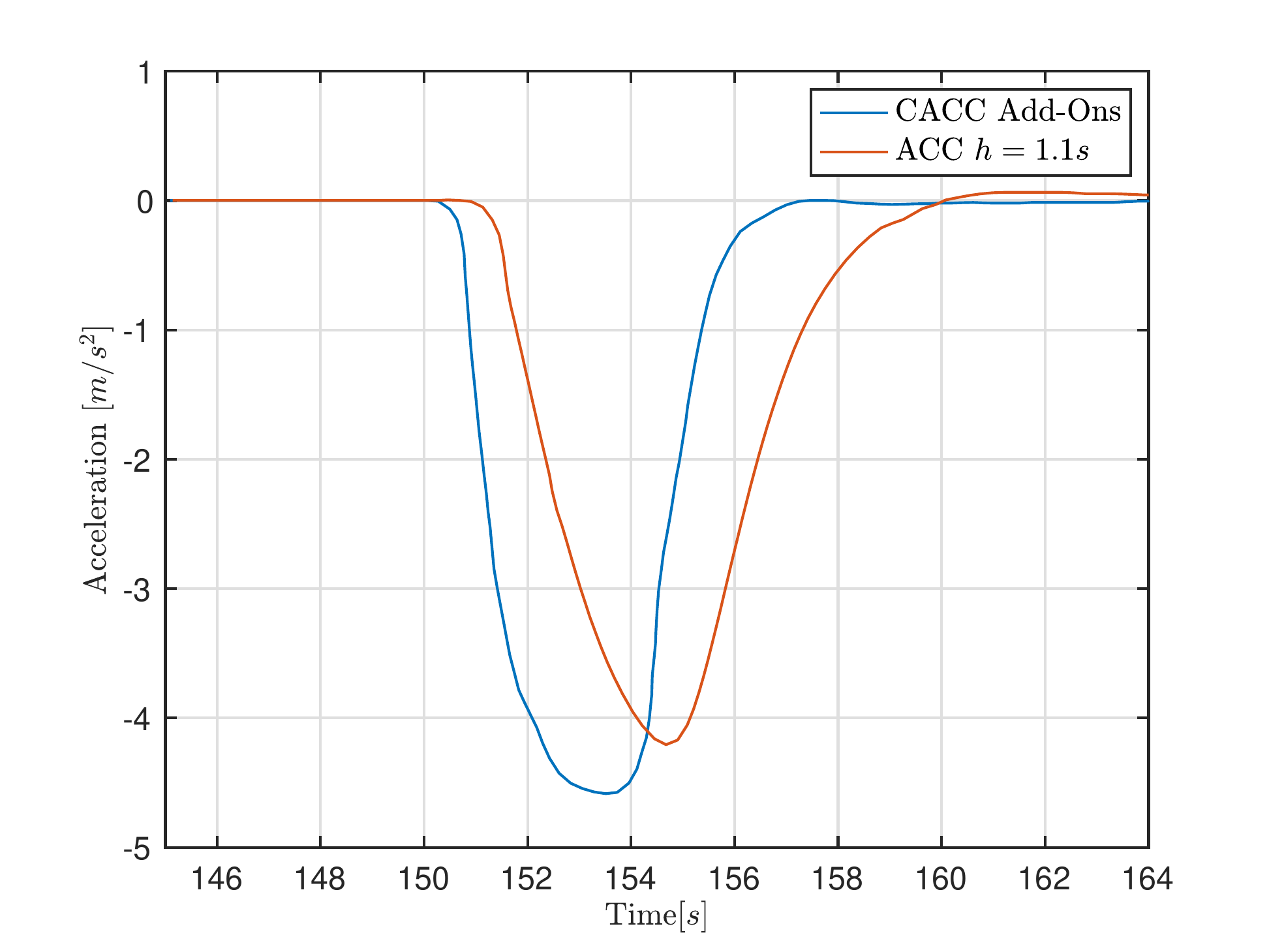}
    \caption{Comparison of the Acceleration profiles of the Ego vehicle for the different control logics}
    \label{accelcomparisonbrak08}
\end{figure}

\noindent From \ref{accelcomparisonbrak08} a comment on the operation of the different logics can be made. 

As expected, the commercial ACC shows a delay, in terms of braking manoeuvre, compared to the connected system. Hence, the CACC system is the one characterized by the highest anticipation of the braking action since it receives the information of real-time acceleration from the vehicles ahead.

\noindent It is important to note that the commercial ACC has to be designed with a higher time gap to have a stable behaviour (for example, $h=1.1s$, as previously done in this work for commercial ACC). Therefore, its delayed response to the breaking manoeuvre and its lower traffic flow capacity is a consequence of this choice.

\begin{figure}
    \centering
    \includegraphics[width=0.8\linewidth]{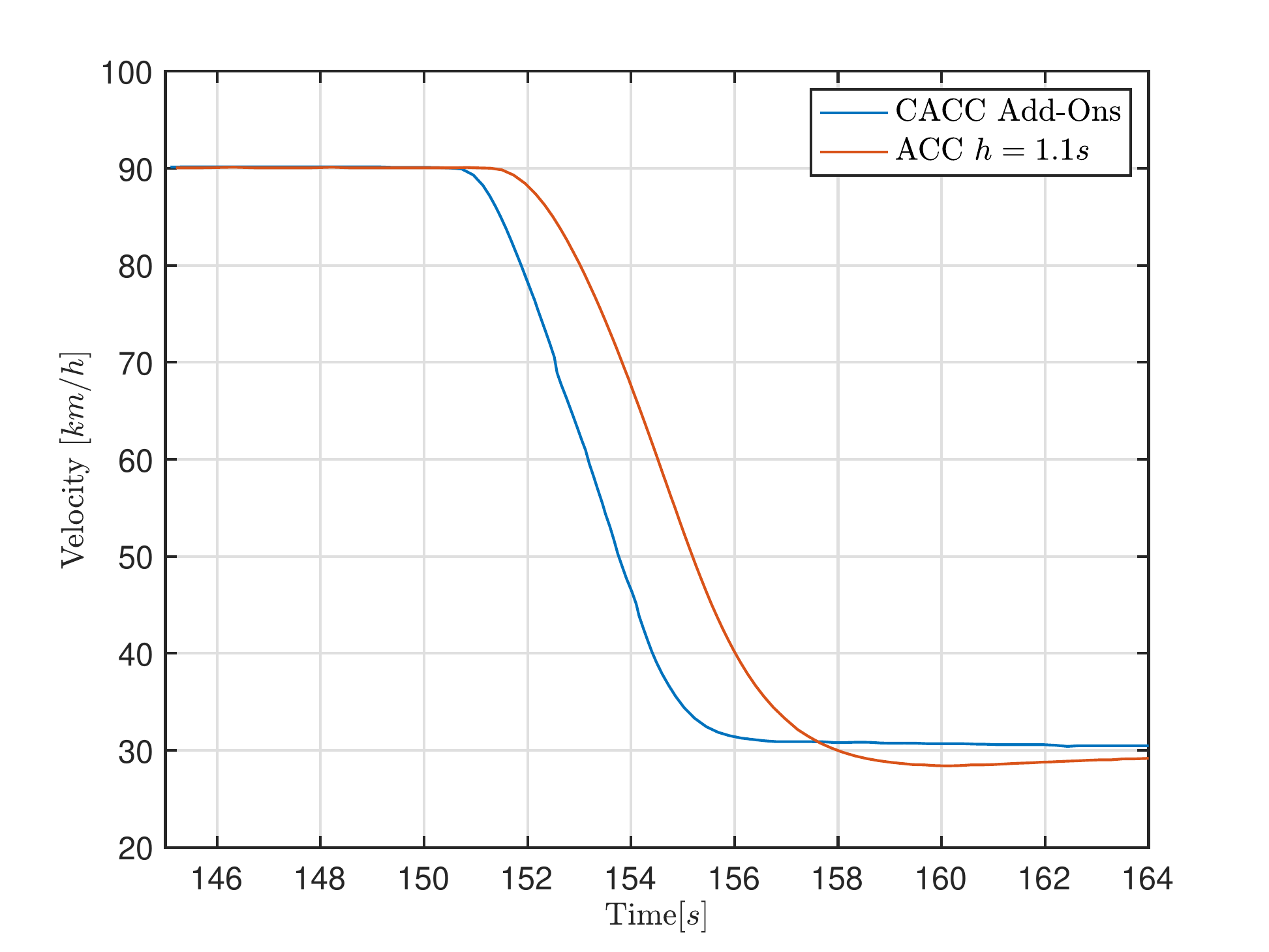}
    \caption{Comparison of the Velocity profiles of the Ego vehicle for the different control logics}
    \label{velcomparisonbrak08}
\end{figure}

\noindent As a result of the previous considerations, the velocity profiles in \ref{velcomparisonbrak08} show once again that the connected systems are able to anticipate the braking manoeuvre.

\begin{figure}
    \centering
    %\captionsetup[subfigure]{justification=centering}
    \includegraphics[width=0.8\linewidth]{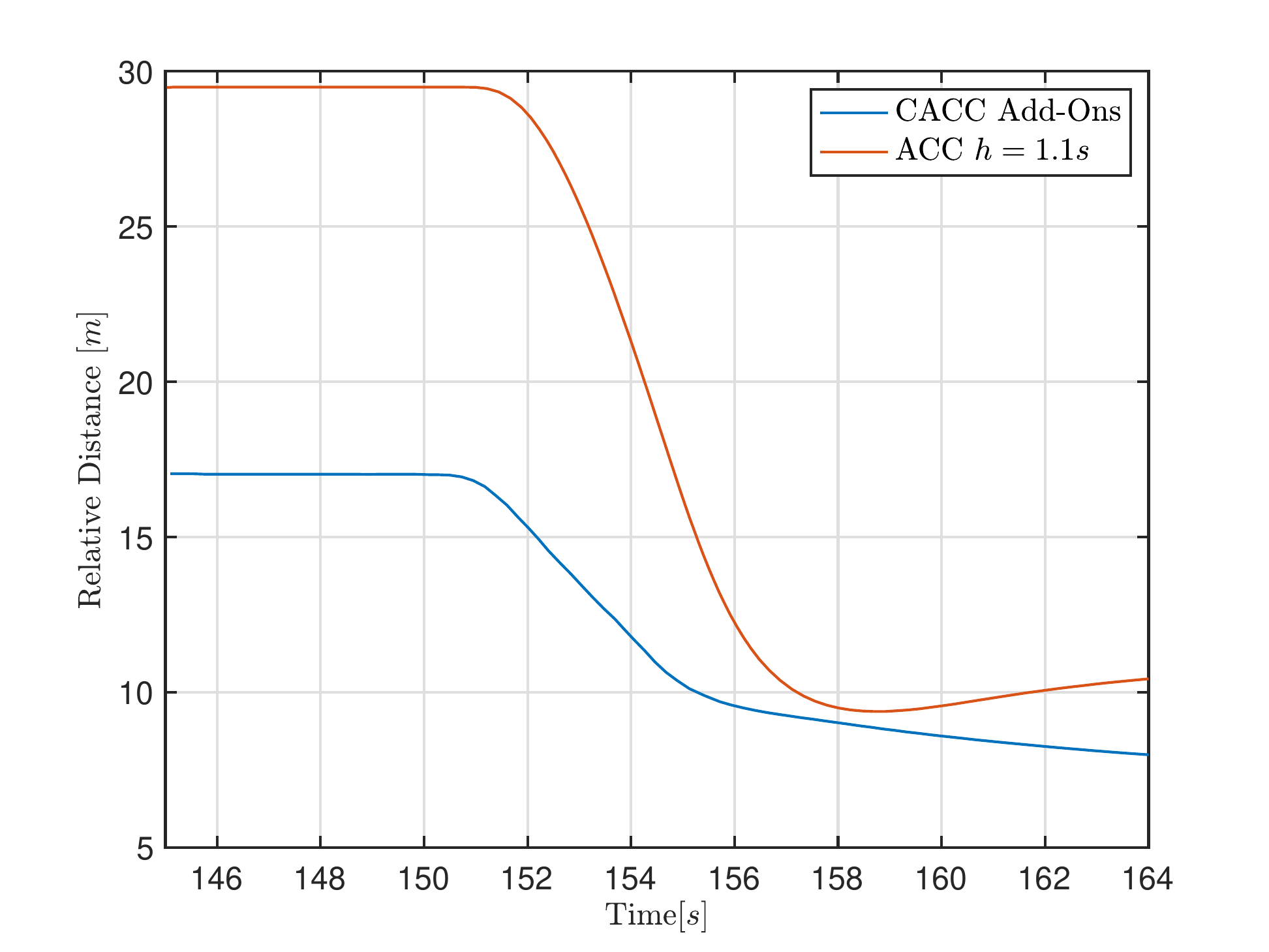}
    \caption{Comparison of the Relative Distances of the Ego vehicle with respect to the Preceding vehicle for the different control logics}
    \label{reldistcomparisonbrak08}
\end{figure}

\noindent It can be seen how the information transmitted by other vehicles to the Ego vehicle is used to guarantee a safer behaviour. 
Looking at \ref{reldistcomparisonbrak08}, advantages that connected systems can bring are well evident: not only safety is guaranteed, but it is also possible to achieve a higher traffic flow capacity since the relative distances are reduced, thus reducing traffic congestion.
 
\begin{figure}[h!]
    \centering
    \hspace{-1cm}
    \includegraphics[width=0.8\linewidth]{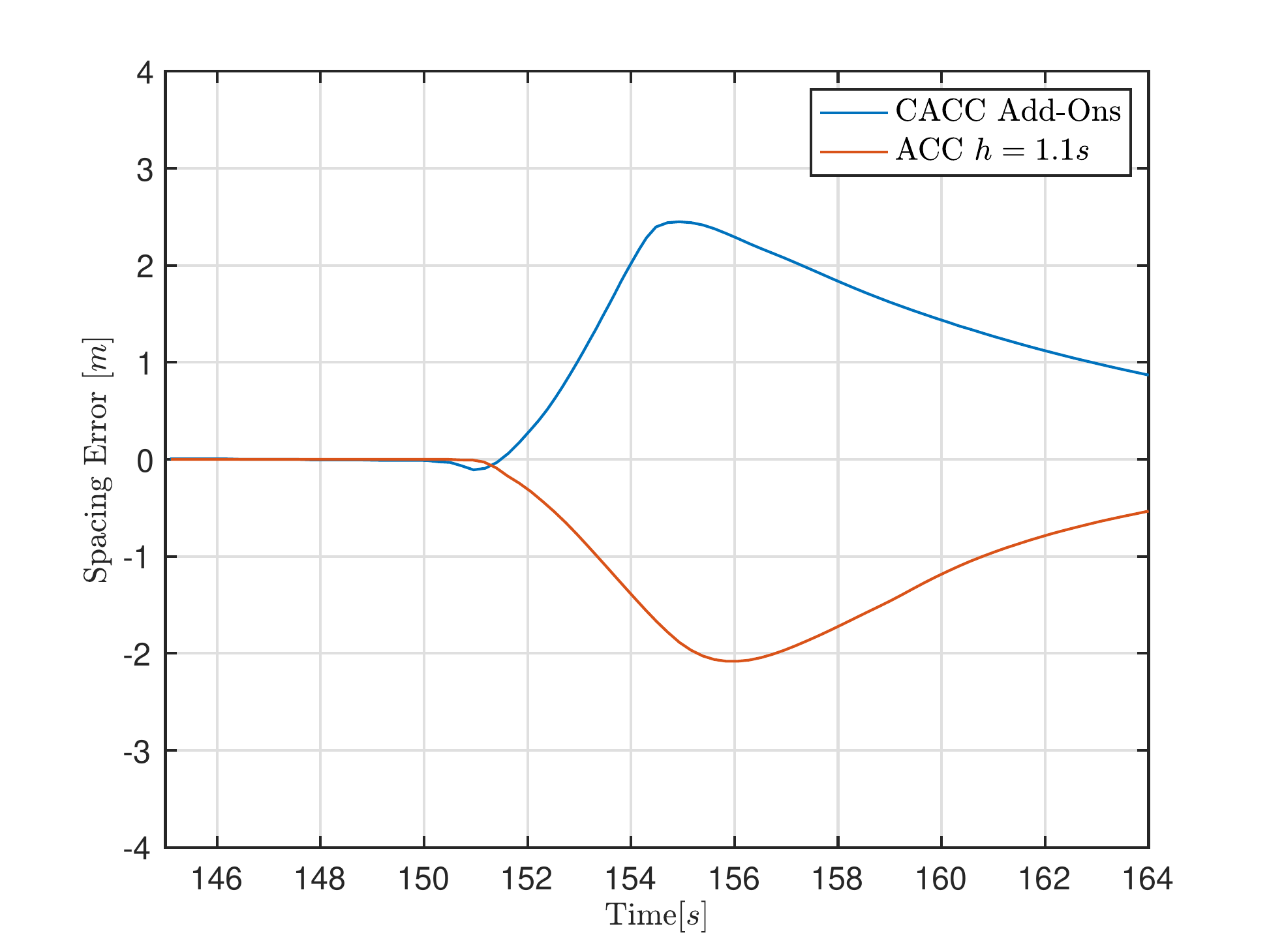}
    \caption{Comparison of the Spacing Errors of the Ego vehicle with respect to the Preceding vehicle for different control logics}
    \label{spacerrcomparisonbrak08}
\end{figure}

By looking at \ref{spacerrcomparisonbrak08}, it is evident how during braking maneuvers the ACC system has a negative spacing error, which means a reduced relative distance between.
Therefore, the ACC system cannot guarantee safe operation due to the significantly low value of the spacing error during the braking manoeuvre. 
For the CACC system, it is important to note that the Ego vehicle keeps a positive spacing error. 

\subsubsection{Low Road Friction}
The tests presented in these sections were conducted considering a low ground friction coefficient set to $\mu_{wet} = 0.5$. 

\begin{figure}
    \centering
    \includegraphics[width=0.8\linewidth]{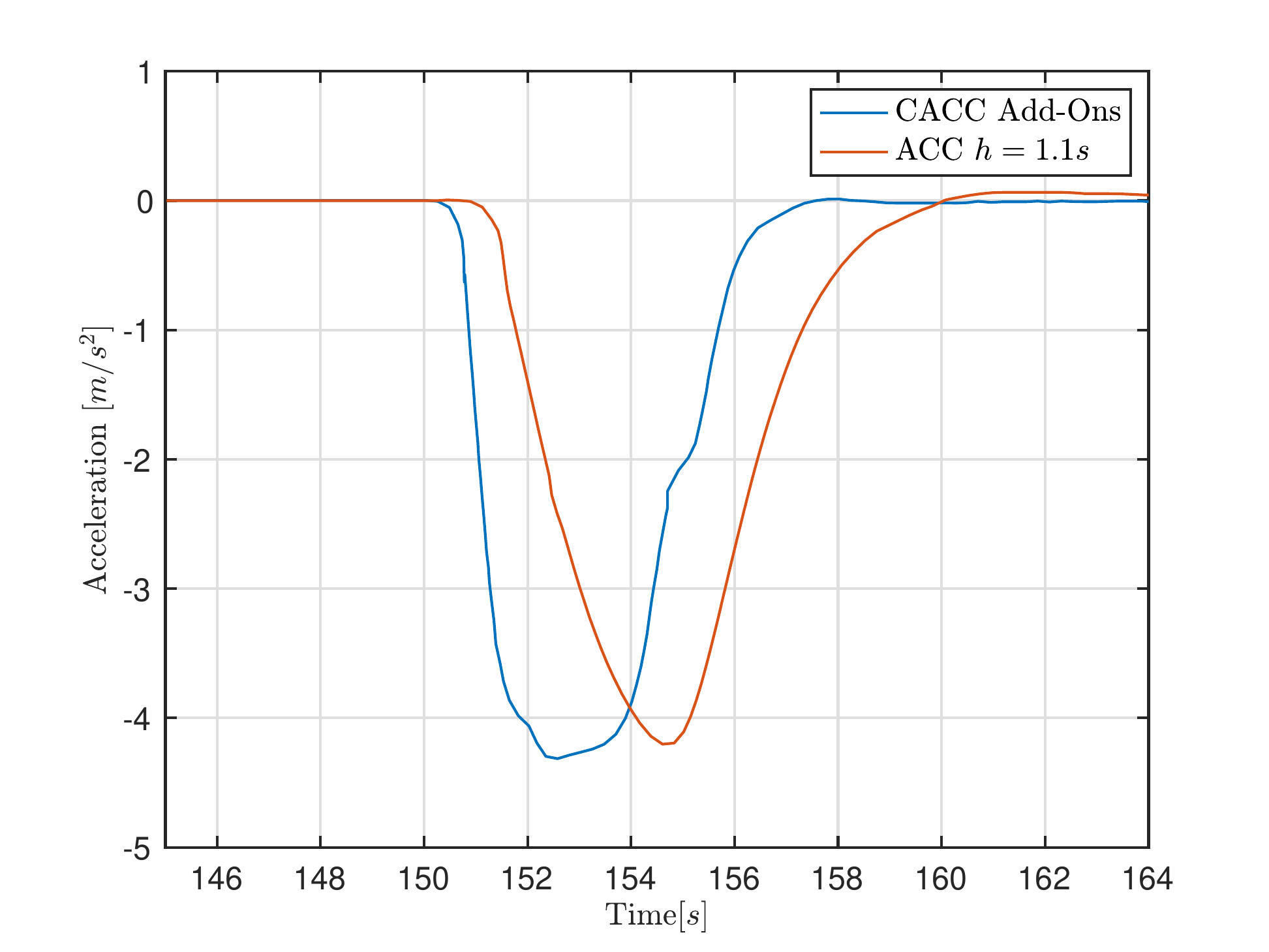}
    \caption{Comparison of the Acceleration profiles of the Ego vehicle for the different control logics}
    \label{accelcomparisonbrak05}
\end{figure}

\noindent Comparing the results shown in \ref{accelcomparisonbrak05} and \ref{accelcomparisonbrak08}, it is possible to highlight one of the limits of the commercial ACC system. The system is not able to modify the behaviour of the vehicle in case of low asphalt adherence because it is not provided with a system that can estimate potential grip. This is clearly not an ideal and safe situation, also by looking at the maximum value of deceleration of the ACC system: almost the entire grip is used in the longitudinal direction thus limiting possible steering actions. Hence, in the case the driver wants to steer in order to avoid a collision, the steering capability can be lower than the case with normal ground friction.\\
In \ref{accelcomparisonbrak05}, the delayed action of the ACC with $h=1.1s$ is visible. However, thanks to a higher time gap value, in this case the system is able to perform a braking maneuver with a comparable level of deceleration with respect to the proposed novel control logics.

\begin{figure}
    \centering
    \includegraphics[width=0.8\linewidth]{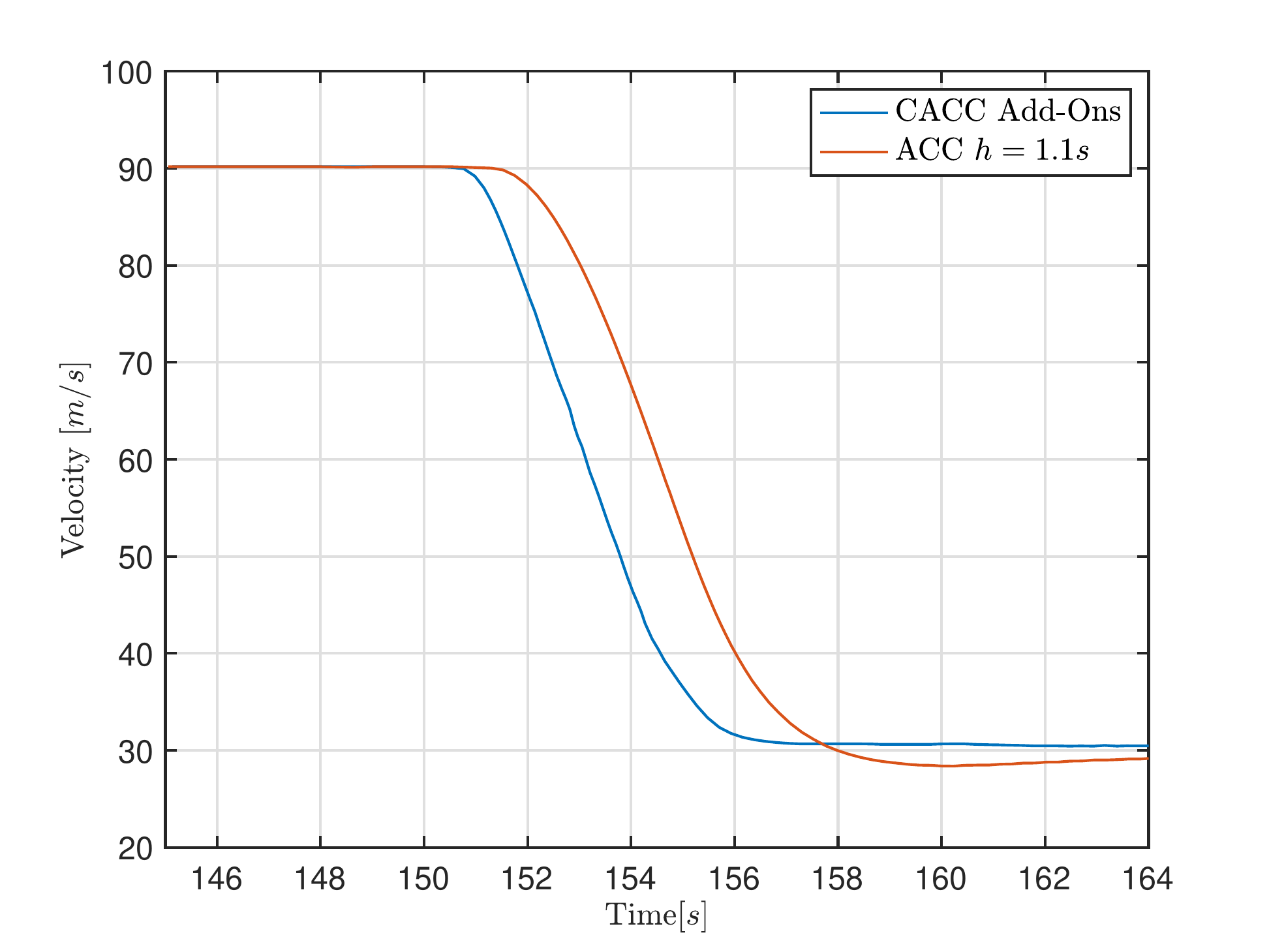}
    \caption{Comparison of the Velocity profiles of the Ego vehicle for the different control logics}
    \label{velcomparisonbrak05}
\end{figure}

\noindent As a result of the acceleration profiles, the velocity profiles in \ref{velcomparisonbrak05} show that the connected systems are able to anticipate the braking manoeuvre.

\begin{figure}
    \centering
    \includegraphics[width=0.76\linewidth]{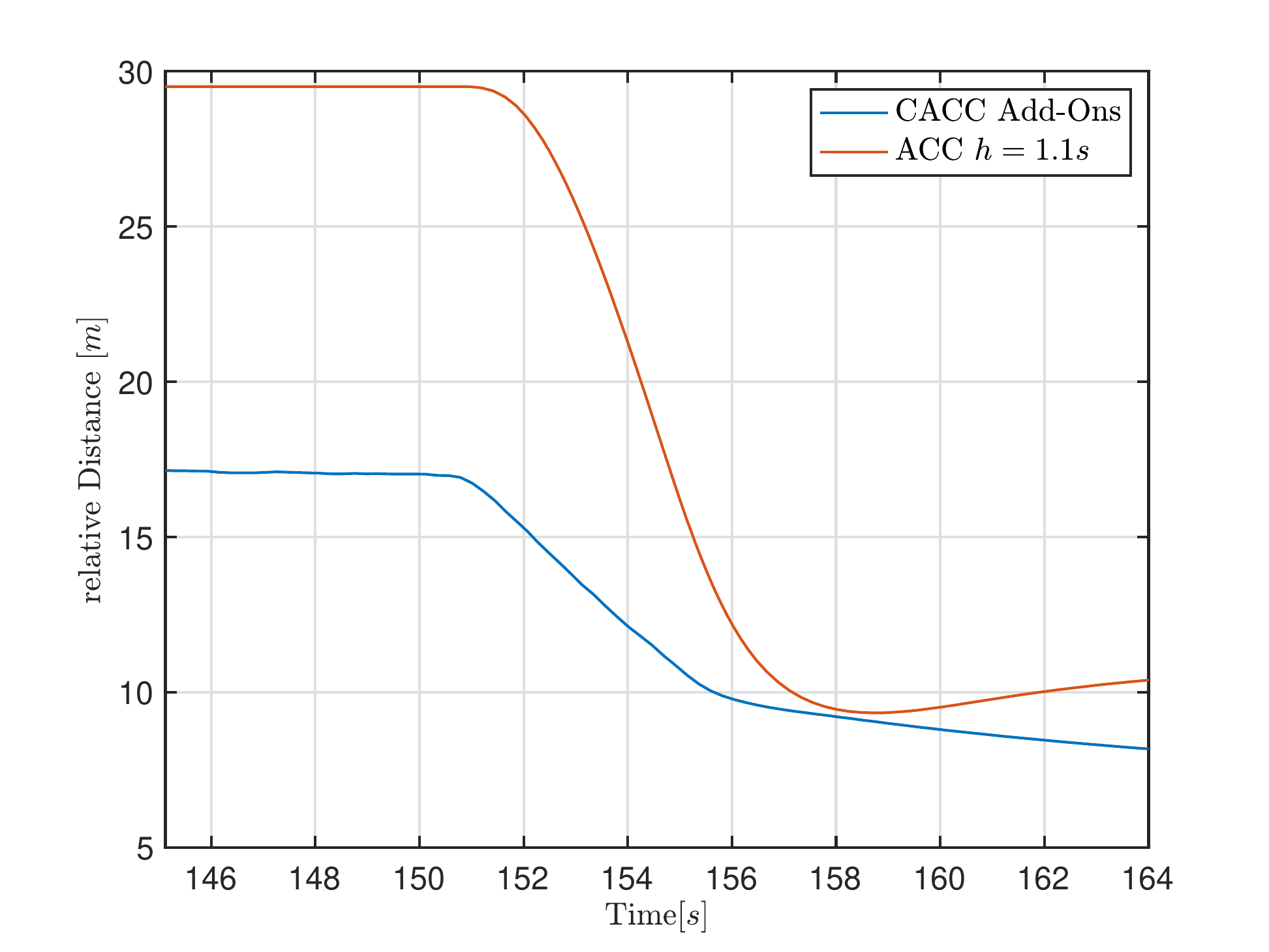}
    \caption{Comparison of the Relative Distances of the Ego vehicle with respect to the Preceding vehicle for the different control logics}
    \label{reldistcomparisonbrak05}
\end{figure}

\noindent For the relative distances, similar comments can be made with respect to the scenario with normal friction.
Looking at \ref{reldistcomparisonbrak05}, the comparison between the novel system and the commercial one with $h = 1.1s$ shows similar minimum values of relative distance, but a higher traffic flow capacity of connected system with respect to the non-connected one.

\begin{figure}
    \centering
    \includegraphics[width=0.8\linewidth]{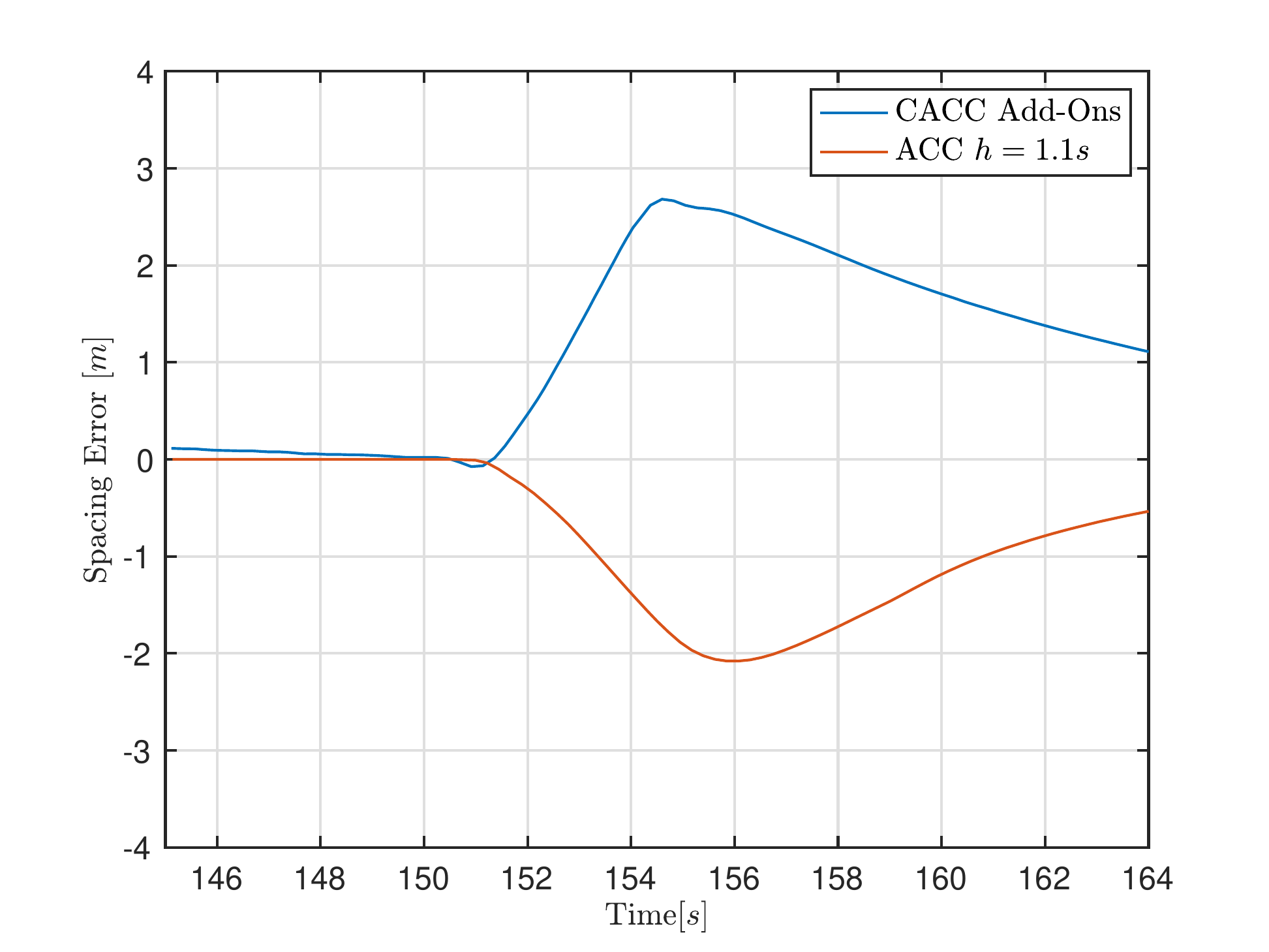}
    \caption{Comparison of the Spacing Errors of the Ego vehicle with respect to the Preceding vehicle for different control logics}
    \label{spacerrcomparisonbrak05}
\end{figure}

From \ref{spacerrcomparisonbrak05}, it can be inferred how the CACC systems promotes safety, allowing the Ego vehicle to keep a positive spacing error with respect to the Preceding vehicle.

In conclusion, the considerations made for the relative distance between vehicles can be further examined considering the traffic flow capacity that is achievable using the different systems. In fact, the Commercial ACC has to adopt a time gap $h = 1.1 s$, while for the connected system, the time gap can be decreased to $h = 0.6 s$. In fact, as already seen, the 5G communication allows shorter inter-vehicles distances without compromising the level of safety. Therefore, it is possible to have a higher traffic flow capacity then with non-connected system, whose time gap has to be kept higher in order to guarantee string stability. In order to quantitatively asses this advantage, the expression for traffic flow capacity $TFC$ introduced in \ref{trafficflow} is used. Its expression is shown in \ref{lane_capac2}:
\begin{equation}
     TFC=\frac{3600v}{L_{v}+L_{c}} \hspace{0.4cm} [vehicles/lane/hours]
     \label{lane_capac2}
\end{equation} 
where:
\begin{itemize}
\setlength\itemsep{0em}
    \item $L_{v}$ is the distance given by the selected control strategy in $[m]$;
    \item $L_{c}$ is the car length in $[m]$;
    \item $v$ is the platoon speed in $[m/s]$.
\end{itemize}

The results for different speed of the leading vehicle are visible in \ref{plot_traf_flow_comp}:
\begin{figure}
    \centering
    \includegraphics[scale=0.55]{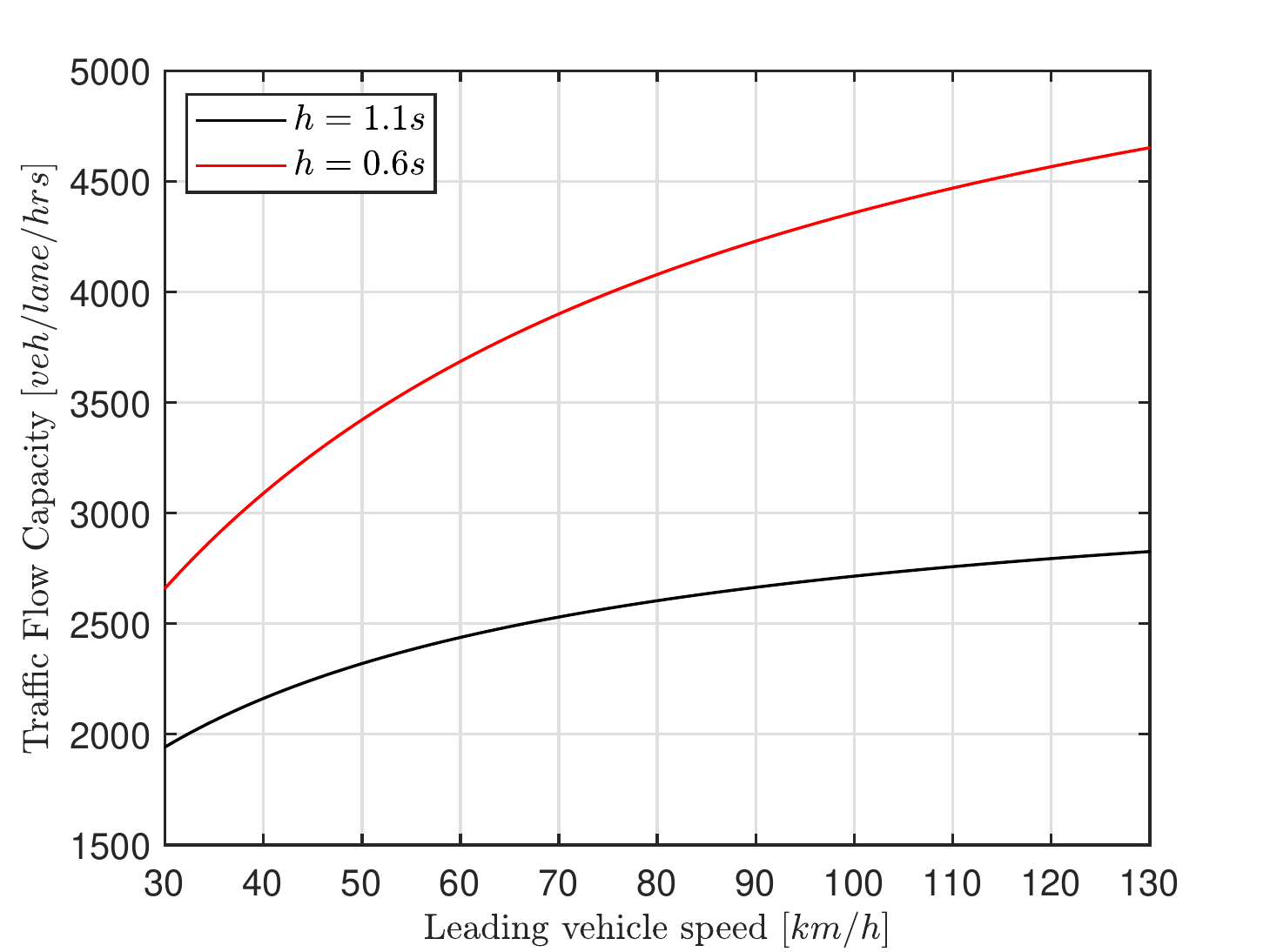}
    \caption{Traffic Flow Capacity for different time gaps}
    \label{plot_traf_flow_comp}
\end{figure}

\noindent The assumptions made for these curves is that the leading vehicles keeps a constant speed. However, in real scenarios there could also be oscillations. As a result, for real situations the values for traffic flow could be also smaller. Nevertheless, it is possible to notice how connected systems can increase considerably the traffic flow due to the lower lower minimum time gap required in order to guarantee safety and string stability.

\subsection{Oscillatory Movement of the Leading Vehicle} \label{oscill_comp_tit}
The comparisons described in this section refer to the scenario described in "Scenario Description". The aim of this section is to show the major differences between the control logics developed in this particular scenario. The following configurations of the systems will be compared:
\begin{itemize}
    \item Commercial ACC system with $h = 1.1 s$;
    \item Connected ACC with Add-Ons with $h = 0.6 s$;
\end{itemize} 
As already said in \ref{scen_oscil}, two different scenarios are presented. For each of them, the comparison of the control logics involves the following parameters:
\begin{itemize}
    \item Ego vehicle acceleration profile;
    \item Ego vehicle velocity profile;
    \item Relative distance between the Ego vehicle and the vehicle ahead.
\end{itemize}
The plots that are shown refer to the behaviour of vehicles after the approaching manoeuvre to the vehicle ahead, so for simulation time between $t=130s$ and $t=210s$. 
Concerning driving comfort, comments are provided based on the behaviour of the acceleration and velocity profiles. The safety features of the control logics can be inferred observing the relative distances and, in particular, their minimum value.
Further comments to verify the strength of the developed control logics in term of comfort can be made considering input for Gas and Brake pedals that the Ego vehicle (i.e., the last vehicle of the string of eight cars) has to apply to follow the desired motion. The virtual testing for the Oscillatory Scenario has been performed assuming that the vehicle model of the vehicles is represented by a First Order Filter, which do not provide Gas and Brake pedals command. 
In order to show the results in term of Gas and Brake pedals, the simulations for the three oscillatory scenarios have been repeated considering that:
\begin{itemize}
    \item For the first seven vehicles, the vehicle model is represented by a First Order Filter;
    \item For the eight vehicles (i.e. the Ego vehicle), the  vehicle model is represented by the CarMaker model.
\end{itemize}
However, before showing the results using the CarMaker model, the following plots are presented:
\begin{enumerate}
    \item  Comparison of Acceleration profiles of the Ego vehicle using firstly the First Order Filter and then the CarMaker model;
    \item Comparison of Relative Distance profiles of the Ego vehicle from the vehicle ahead using firstly the First Order Filter and then the CarMaker model.
\end{enumerate}
Each of these plots is presented for each of the control logics used for the comparison (i.e., ACC, CACC with Add-Ons). The reason why also these plots are shown is that, in this way, it is possible to demonstrate that the comments made so far concerning string stability are not influenced by the use of the First Order Filter instead of the CarMaker model. In fact, the CarMaker model has proven to present a faster response in case of a step input and also to be quicker in following an oscillatory input. However, a verification concerning string stability was left open and it is dealt with in this chapter.

\subsubsection{Oscillatory Scenario No.1}
\begin{figure}
     %\centering
     %\begin{adjustwidth}{-1.5cm}{-1.5cm}
%\captionsetup[subfigure]{justification=centering}
    \centering
    \includegraphics[width=0.76\linewidth]{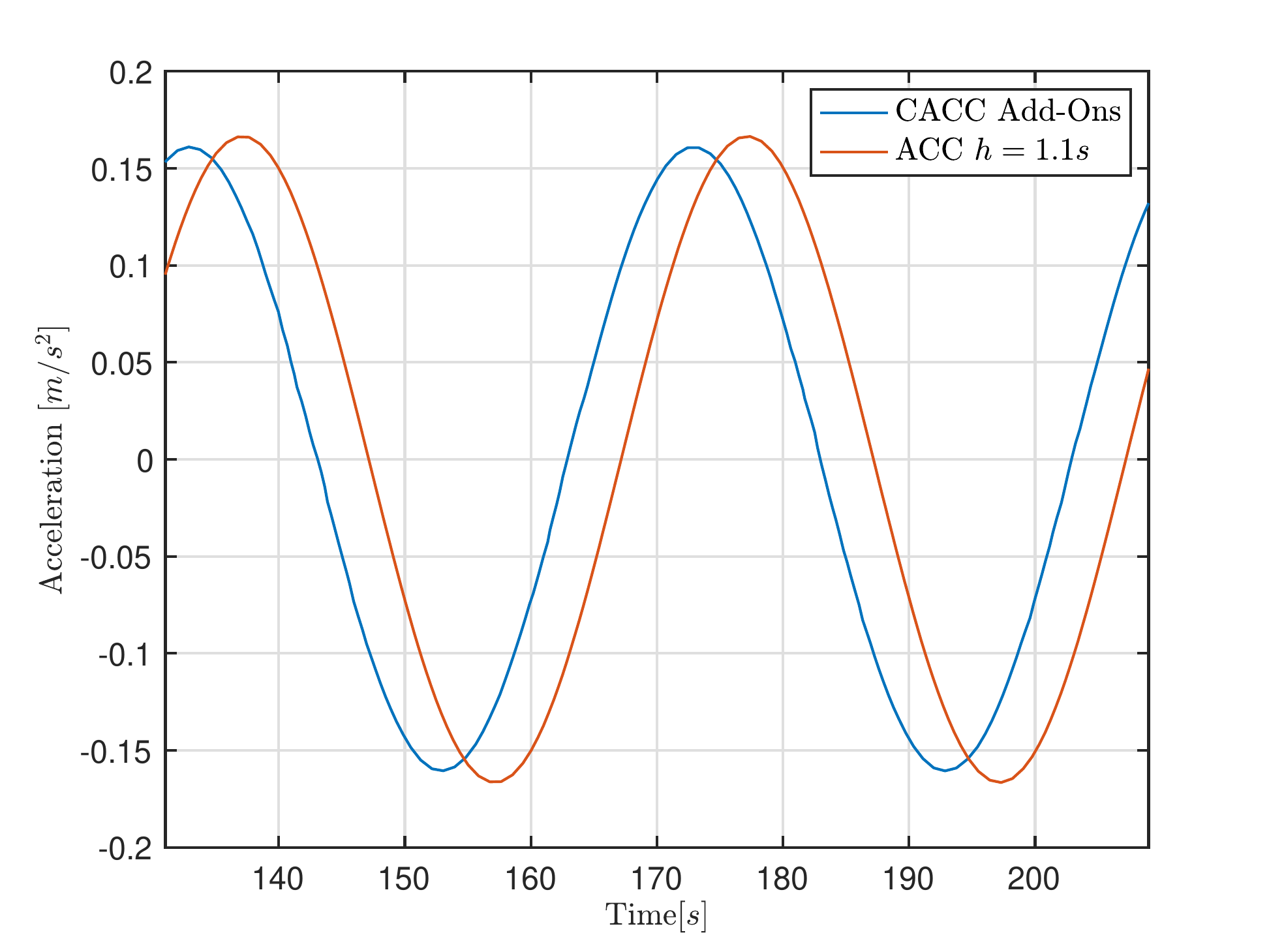}
    \caption{Comparison of the Acceleration profile of the Ego vehicle for different control logics}
    \label{accelcomparisonoscill1}
\end{figure}

As described in \ref{scen_oscil}, the leading vehicle has the following imposed velocity profile:
\begin{itemize}
    \item $v_{lead}=80\pm4km/h$ 
    \item $T=40s$
\end{itemize}
The advantages of using a connected system is represented by the possibility of increasing the traffic flow capacity still guaranteeing a better level of comfort and safety with respect to the Commercial ACC, as the acceleration profiles in \ref{accelcomparisonoscill1} show. The information coming from the "Leading" vehicle of the selected Information Flow Topology (IFT) are used to damp the acceleration oscillations. In fact, if an ACC system with a time gap $h=0.6s$ is used, the string instability associated to the selection of a too low time gap would make the oscillations amplify even more. The most remarkable improvement can be proved by the  10\% reduction of the amplitude of acceleration oscillations with respect to the Commercial ACC. Concerning comfort, the values for RMS of the last car of the vehicles string reflect the described situation and are reported in \ref{RMS_oscill1}:
\begin{table}
\footnotesize
    \centering
     \begin{tabular}{c c c} 
     \hline
     Control Logic & Time Gap [$s$] & RMS of Acceleration [$m/s^2$]\\
     \hline
     
     Commercial ACC & $h=1.1s$ & $0.121$ \\ 
     
     Connected ACC with Add-Ons & $h=0.6s$ & $0.101$ \\

     \hline
    \end{tabular}
    \caption{RMS of Acceleration Profile for Last Vehicle of the String}
    \label{RMS_oscill1}
\end{table}

\noindent 
Observing the behaviour of the relative distances profile in \ref{reldistcomparisonoscill1}, it is visible how the ACC system ensures higher inter-vehicle distances, at the cost of a reduced traffic capacity. The latter can be increased using connected systems, even if shorter inter-vehicle distances might be a problem in term of safety. However, as already proved before, connected systems have produced better results then a commercial ACC with the same time gap.

\begin{figure}
    \centering
    \includegraphics[width=0.76\linewidth]{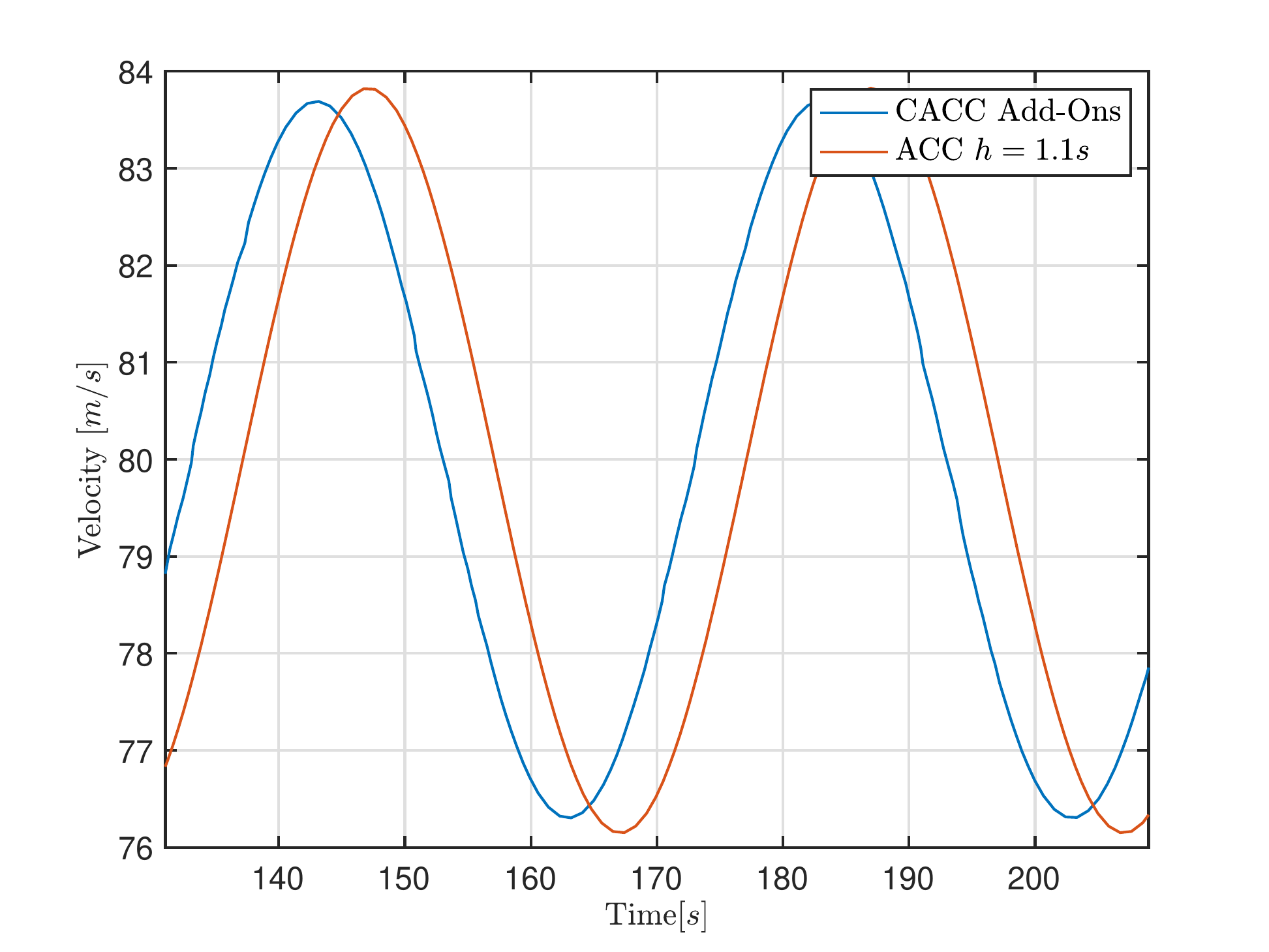}
    \caption{Comparison of the Velocity profile of the Ego vehicle for different control logics}
    \label{velcomparisonoscill1}
\end{figure}

\begin{figure}
    \centering
    \includegraphics[width=0.76\linewidth]{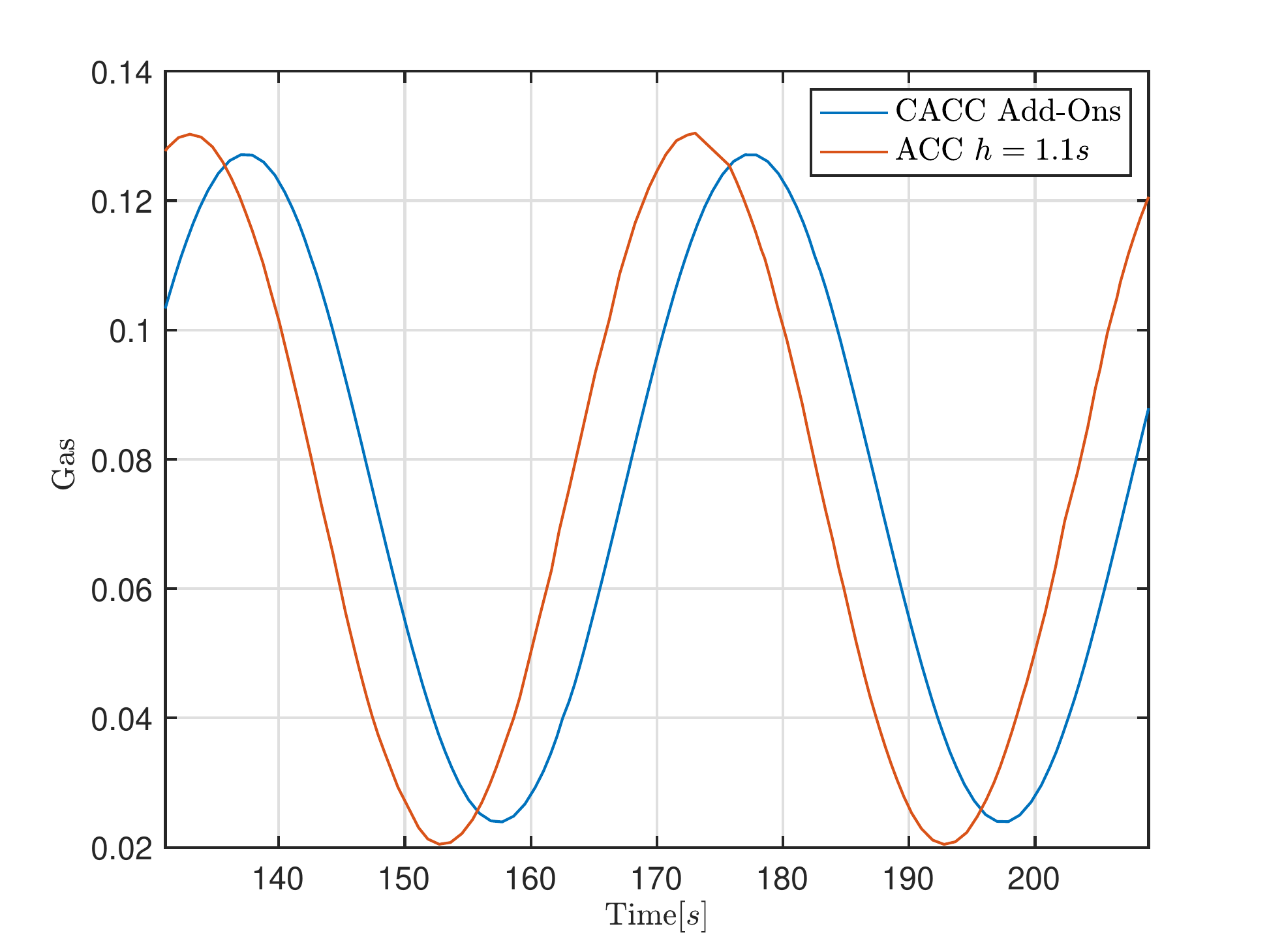}
    \caption{Gas and Brake Pedal for Ego Vehicle for different control logics}
    \label{gascomparisonoscill1}
\end{figure}

Comparing the graphs shown in \ref{gascomparisonoscill1}, for each control logic the deceleration phase is accomplished simply reducing the Gas input and not operating on the Brake. Both the connected systems are able to reduce the Gas pedal compared to the commercial ACC. Furthermore, it is visible how the smoother profile can help the vehicle in guaranteeing a more comfortable drive. Finally, these consideration can intuitively have an impact on the fuel consumption: considering the curves in \ref{gascomparisonoscill1} and the previous work of Li et al. \cite{li2010model}, reduction on accelerations (directly related to a lower gas pedal percentage) have a similar effect of reduction on fuel consumption.

\subsubsection{Oscillatory Scenario No.2}
\begin{figure}
    \centering
    \includegraphics[width=0.76\linewidth]{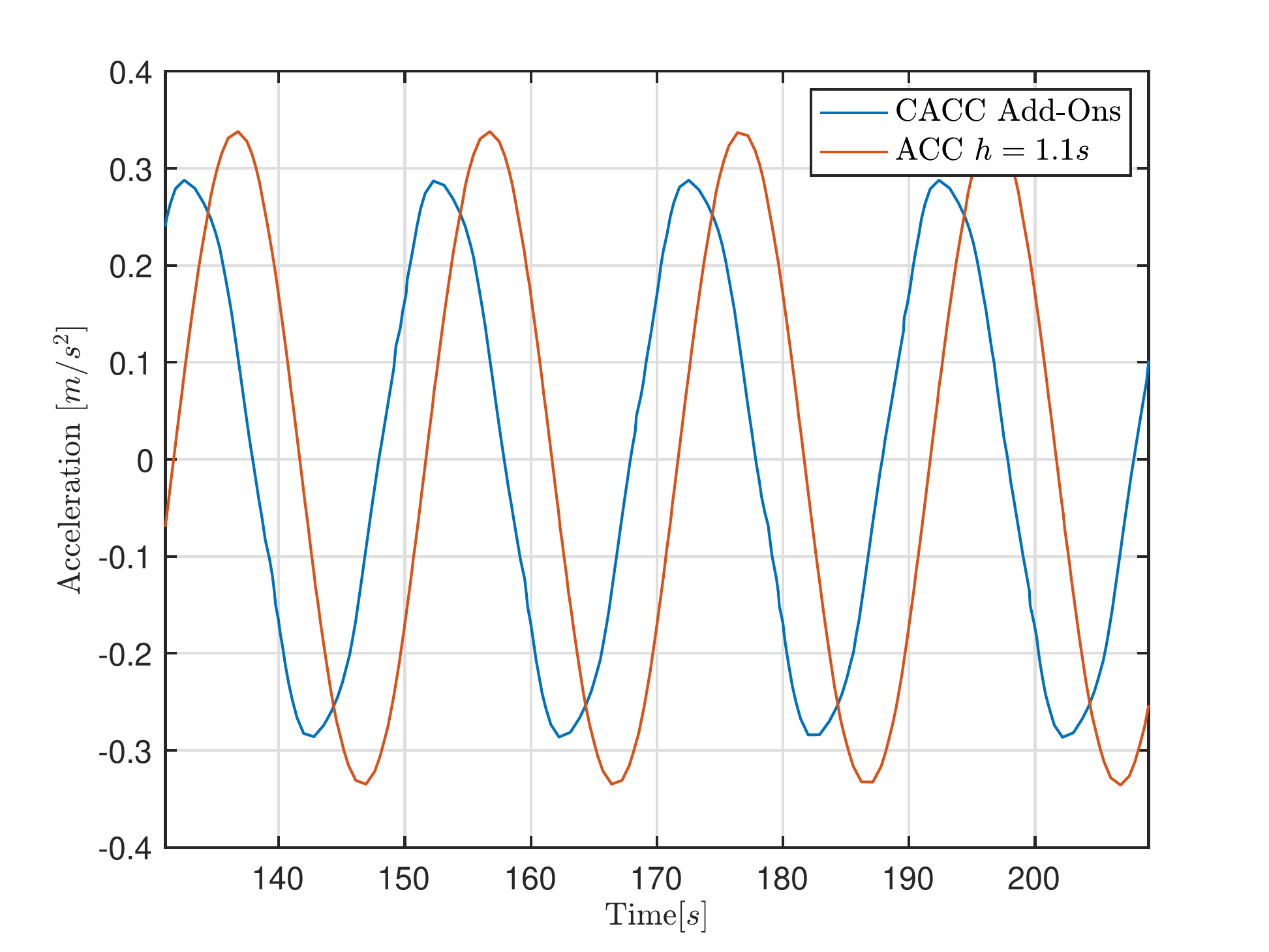}
    \caption{Comparison of the Acceleration profile of the Ego vehicle for different control logics}
    \label{accelcomparisonoscill3}
\end{figure}

As described in \ref{scen_oscil}, the leading vehicle has the following imposed velocity profile:
\begin{itemize}
    \item $v_{lead}=80\pm4km/h$ 
    \item $T=20s$
\end{itemize}
As shown in \ref{accelcomparisonoscill3}, it is possible to notice that the amplitude of the oscillations considering the CACC with Add-Ons approach is reduced up to almost 20\% with respect to the ACC systems. This great difference can be attributed mainly to the fact that the information coming from the vehicles ahead in the line are used to reduce the oscillations of acceleration.

Considering the acceleration profile of Commercial ACC instead, in order to avoid risk of collisions, the time gap must be increased and therefore traffic capacity is decreased. Concerning comfort, the advantages of a connected ACC are evident, also looking at the values for RMS of the last car of the vehicle string reported in \ref{RMS_oscill3}

\begin{table}[h!]
    \centering
     \begin{tabular}{c c c} 
     \hline
     Control Logic & Time Gap [$s$] & RMS of Acceleration [$m/s^2$]\\
     \hline
     
     Commercial ACC & $h=1.1s$ & $0.248$ \\ 
     
     Connected ACC with Add-Ons & $h=0.6s$ & $0.170$ \\

     \hline
    \end{tabular}
    \caption{RMS of Acceleration Profile for Last Vehicle of the String}
    \label{RMS_oscill3}
\end{table}

\begin{figure}
    \centering
    \includegraphics[width=0.76\linewidth]{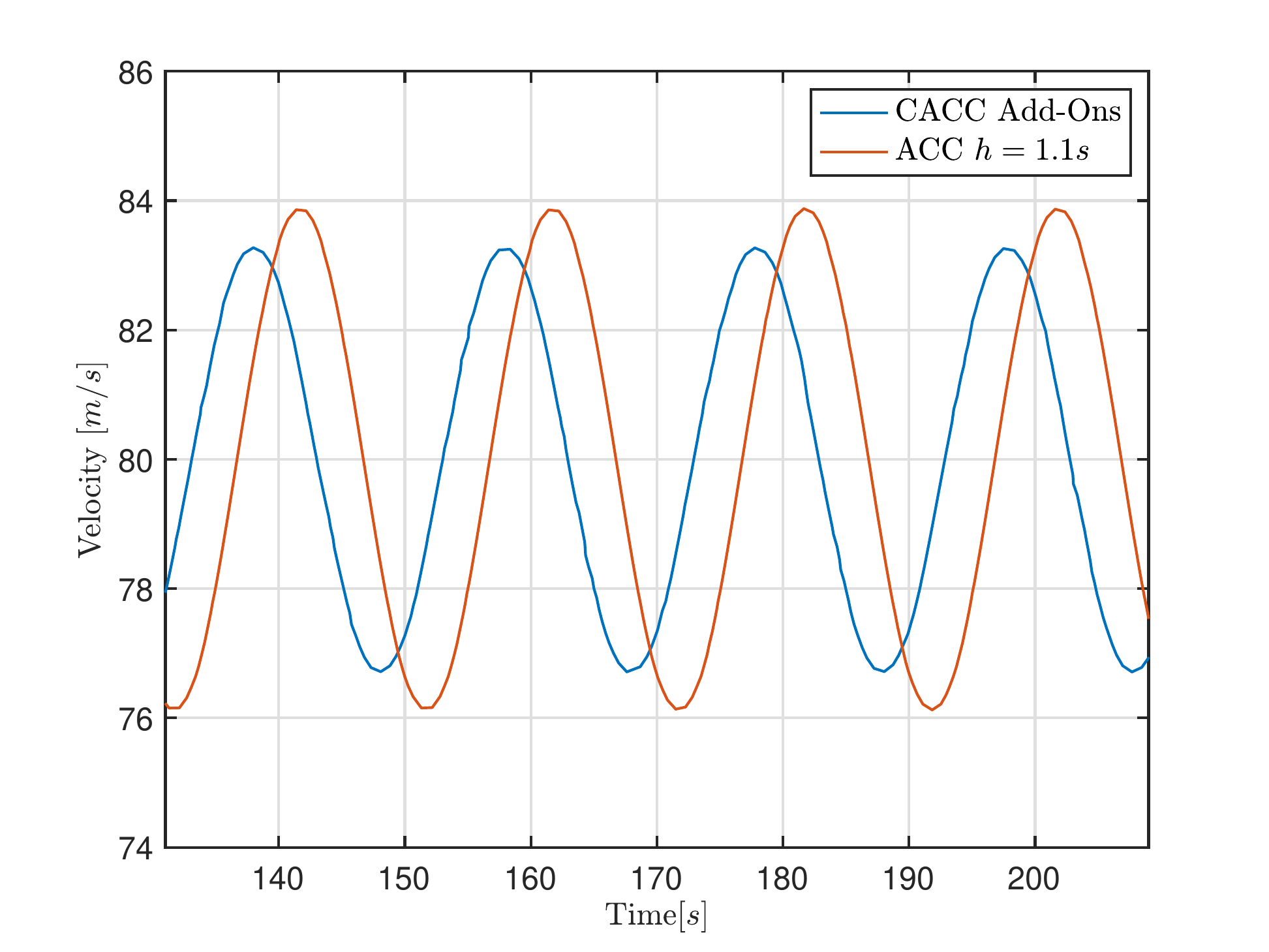}
    \caption{Comparison of the Velocity profile of the Ego vehicle for different control logics}
    \label{velcomparisonoscill3}
\end{figure}

As a result of this, also the velocity profiles shown in \ref{velcomparisonoscill3} show that connected system presents a reduced amplitude and confirms that the system based on the connected controller is able follow the string of vehicles with a velocity oscillation $1km/h$ less wide then the system with Add-Ons.

Considering \ref{gascomparisonoscill2}, it is evident how the connected system reduce amplitude of deceleration with respect to a commercial ACC. Furthermore, it is important to remark that also the Gas pedal presents a lower amplitude for the connected systems with respect to a Commercial ACC. The CACC with Add-ons shows a 15\% reduction of the Gas input with respect of the commercial system and a 5\% reduction the Brake pedal input .\\
As a result of this, the driving results more comfortable since no unnecessary acceleration and deceleration are needed. Improvements are expected also considering the fuel consumption.

\begin{figure}
    \centering
    \includegraphics[width=0.76\linewidth]{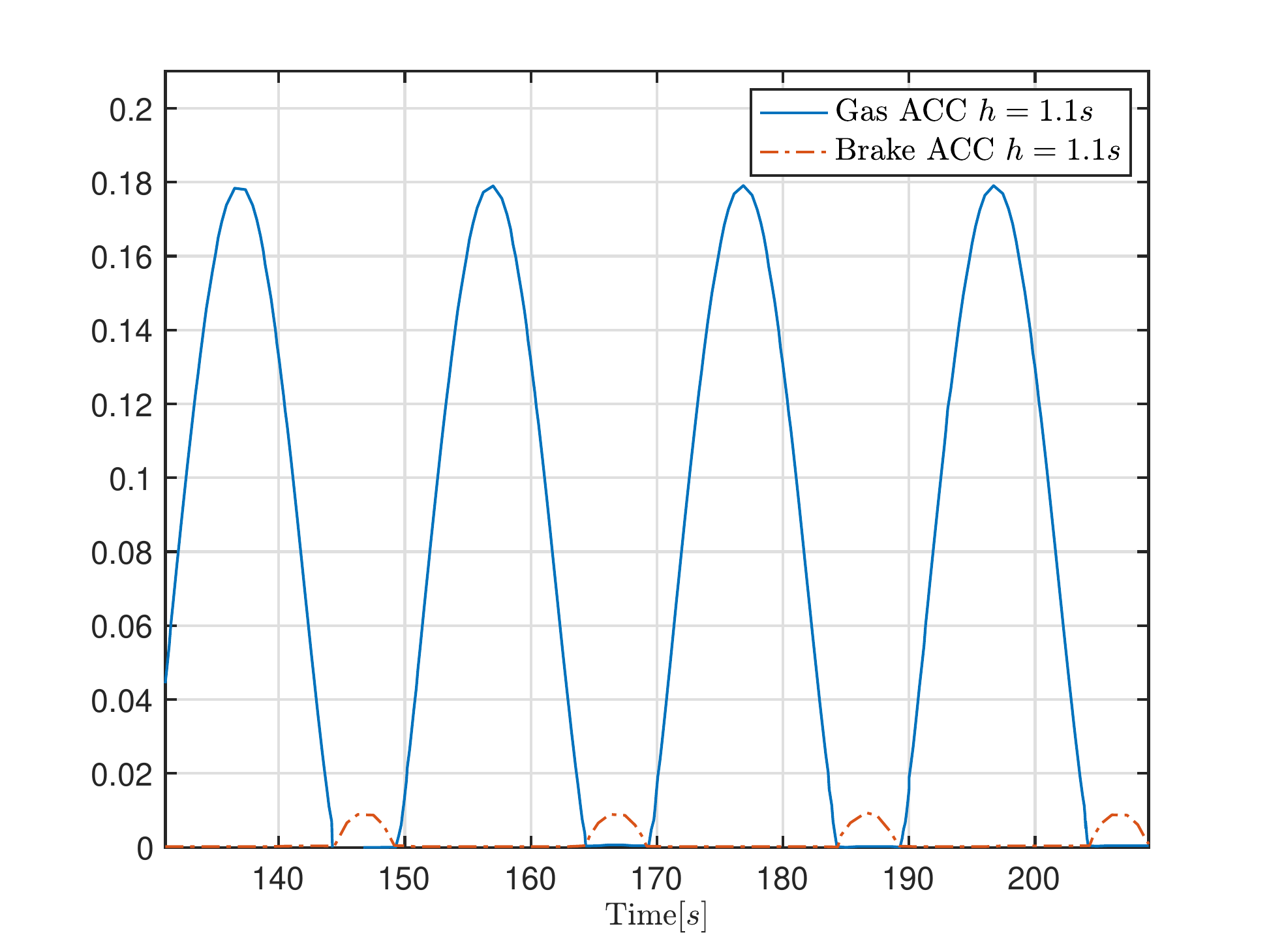}
    \caption{Gas and Brake Pedal for Ego Vehicle for ACC system}
    \label{gascomparisonoscill2}
\end{figure}

\begin{figure}
    \centering
    \includegraphics[width=0.76\linewidth]{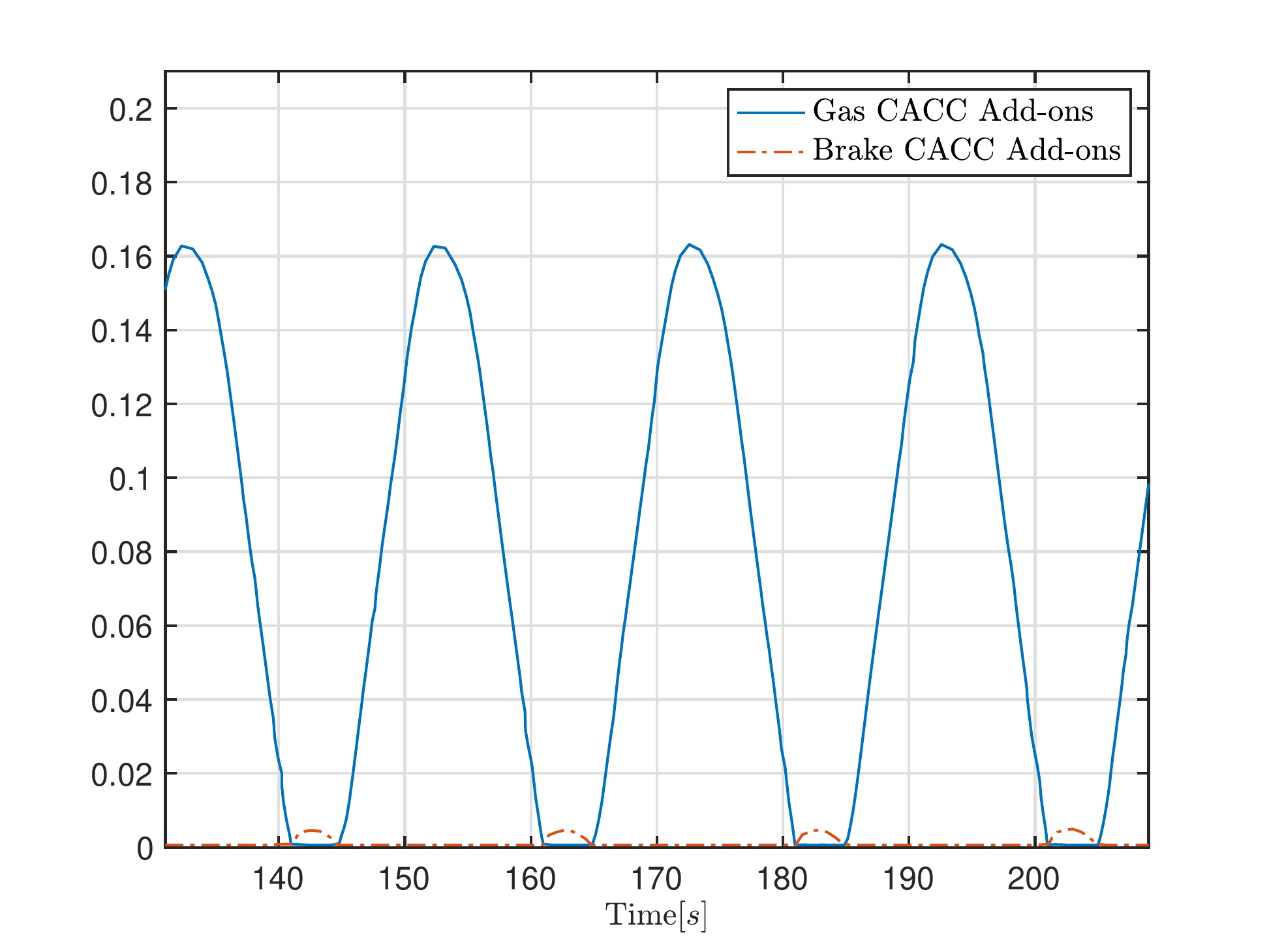}
    \caption{Gas and Brake Pedal for Ego Vehicle for CACC with Add-Ons system}
    \label{gascomparisonoscill2}
\end{figure}

\section{Conclusion}
In conclusion, as defined in \ref{intro}, the aim of the work was showing how connectivity between vehicles and infrastructures can increment road safety and comfort, as well as increase traffic capacity if properly integrated in ADAS algorithm development
The targets can be summarized by the following tasks: 
\begin{itemize}
    \item Show the convenience of connectivity by focusing on the use of fast and reliable 5G communication for Connected ACC systems;
    \item Investigate the benefits of using information about potential grip from Pirelli Cyber Tyre in the design of connected systems;
\end{itemize}
After an accurate literature review about ACC and CACC logics in order to highlight advantages of connectivity applied to commercial ACC systems, at first a commercial ACC based on the most common techniques adopted in literature has been developed. Then, a \textit{novel} control approach was proposed. 

The\textit{novel} approach was developed by considering the assumption of maintaining an architecture as close as possible to the one of the partner companies involved in the project. In this way, the resulting system is easier and faster to be implemented on a commercial vehicle (since the main architecture and control boards are maintained) and with limited costs. Therefore, the CACC with Add-Ons system was designed.
\\
Considering the results obtained in the virtual testing, it can be stated that the use of 5G communication allows to overcome the limits of the commercial ACC systems. In particular, the braking manoeuvre is anticipated with respect to the movement of the vehicles ahead and a higher inter-vehicle distance during the manoeuvre is ensured. Thus, it is possible to increase the traffic flow capacity, guaranteeing string stability. At the same time, the use of Cyber Tyre permits a reduction of deceleration value in case of low adherence.
 \bibliographystyle{elsarticle-num}





% \end{thebibliography}
\end{document}